# Molecular and Serologic Diagnostic Technologies for SARS-CoV-2

# Authors

- **Halie M. Rando** 0000-0001-7688-1770 rando2 tamefoxtime  Department of Systems Pharmacology and Translational Therapeutics, University of Pennsylvania, Philadelphia, Pennsylvania, United States of America; Department of Biochemistry and Molecular Genetics, University of Colorado School of Medicine, Aurora, Colorado, United States of America; Center for Health AI, University of Colorado School of Medicine, Aurora, Colorado, United States of America · Funded by the Gordon and Betty Moore Foundation (GBMF 4552); the National Human Genome Research Institute (R01 HG010067)

- **Christian Brueffer** 0000-0002-3826-0989 cbrueffer cbrueffer  Department of Clinical Sciences, Lund University, Lund, Sweden

- **Ronan Lordan** 0000-0001-9668-3368 RLordan el_ronan  Institute for Translational Medicine and Therapeutics, Perelman School of Medicine, University of Pennsylvania, Philadelphia, PA 19104-5158, USA; Department of Medicine, Perelman School of Medicine, University of Pennsylvania, Philadelphia, PA 19104, USA; Department of Systems Pharmacology and Translational Therapeutics, Perelman School of Medicine, University of Pennsylvania; Philadelphia, PA 19104, USA

- **Anna Ada Dattoli** 0000-0003-1462-831X aadattoli aadattoli  Department of Pathology and Laboratory Medicine, The Children's Hospital of Philadelphia, Philadelphia, PA, USA; Department of Systems Pharmacology & Translational Therapeutics, Perelman School of Medicine, University of Pennsylvania, Philadelphia, PA, USA

- **David Manheim** 0000-0001-8599-8380 davidmanheim davidmanheim  1DaySooner, Delaware, United States of America; Risk and Health Communication Research Center, School of Public Health, University of Haifa, Haifa, Israel; Technion, Israel Institute of Technology, Haifa, Israel · Funded by Center for Effective Altruism, Long Term Future Fund

- **Jesse G. Meyer** 0000-0003-2753-3926 jessegmeyerlab  Department of Biochemistry, Medical College of Wisconsin, Milwaukee, Wisconsin, United States of America · Funded by National Institute of General Medical Sciences (R35 GM142502)

- **Ariel I. Mundo** 0000-0002-6014-4538 aimundo Department of Biomedical Engineering, University of Arkansas, Fayetteville, Arkansas, USA

- **Dimitri Perrin** 0000-0002-4007-5256 SystemsResearch dperrin School of Computer Science, Queensland University of Technology, Brisbane, Australia; Centre for Data Science, Queensland University of Technology, Brisbane, Australia

- **David Mai** 0000-0002-9238-0164 davemai lococyte Department of Bioengineering, University of Pennsylvania, Philadelphia, PA, USA; Center for Cellular Immunotherapies, Perelman School of Medicine, and Parker Institute for Cancer Immunotherapy at University of Pennsylvania, Philadelphia, PA, USA

- **Nils Wellhausen** 0000-0001-8955-7582 nilswellhausen Department of Systems Pharmacology and Translational Therapeutics, University of Pennsylvania, Philadelphia, Pennsylvania, United States of America

- **COVID-19 Review Consortium**

- **Anthony Gitter** 0000-0002-5324-9833 agitter anthonygitter Department of Biostatistics and Medical Informatics, University of Wisconsin-Madison, Madison, Wisconsin, United States of America; Morgridge Institute for Research, Madison, Wisconsin, United States of America · Funded by John W. and Jeanne M. Rowe Center for Research in Virology

- **Casey S. Greene** 0000-0001-8713-9213 cgreene GreeneScientist Department of Systems Pharmacology and Translational Therapeutics, University of Pennsylvania, Philadelphia, Pennsylvania, United States of America; Childhood Cancer Data Lab, Alex's Lemonade Stand Foundation, Philadelphia, Pennsylvania, United States of America; Department of Biochemistry and Molecular Genetics, University of Colorado School of Medicine, Aurora, Colorado, United States of America; Center for Health AI, University of Colorado School of Medicine, Aurora, Colorado, United States of America · Funded by the Gordon and Betty Moore Foundation (GBMF 4552); the National Human Genome Research Institute (R01 HG010067)

Authors with similar contributions are ordered alphabetically.

**COVID-19 Review Consortium:** Vikas Bansal, John P. Barton, Simina M. Boca, Joel D Boerckel, Christian Brueffer, James Brian Byrd, Stephen Capone, Shikta Das, Anna Ada Dattoli, John J. Dziak, Jeffrey M. Field, Soumita Ghosh, Anthony Gitter, Rishi Raj Goel, Casey S. Greene, Marouen Ben Guebila, Daniel S. Himmelstein, Fengling Hu, Nafisa M. Jadavji, Jeremy P. Kamil, Sergey Knyazev, Likhitha Kolla, Alexandra J. Lee, Ronan Lordan, Tiago Lubiana, Temitayo Lukan, Adam L. MacLean, David Mai, Serghei Mangul, David Manheim, Lucy D'Agostino McGowan, Jesse G. Meyer, Ariel I. Mundo, Amruta Naik, YoSon Park, Dimitri Perrin, Yanjun Qi, Diane N. Rafizadeh, Bharath Ramsundar, Halie M. Rando, Sandipan Ray, Michael P. Robson, Vincent Rubinetti, Elizabeth Sell, Lamonica Shinholster, Ashwin N. Skelly, Yuchen Sun, Yusha Sun, Gregory L Szeto, Ryan Velazquez, Jinhui Wang, Nils Wellhausen

# 1 Abstract

The COVID-19 pandemic has presented many challenges that have spurred biotechnological research to address specific problems. Diagnostics is one area where biotechnology has been critical. Diagnostic tests play a vital role in managing a viral threat by facilitating the detection of infected and/or recovered individuals. From the perspective of what information is provided, these tests fall into two major categories, molecular and serological. Molecular diagnostic techniques assay whether a virus is present in a biological sample, thus making it possible to identify individuals who are currently infected. Additionally, when the immune system is exposed to a virus, it responds by producing antibodies specific to the virus. Serological tests make it possible to identify individuals who have mounted an immune response to a virus of interest and therefore facilitate the identification of individuals who have previously encountered the virus. These two categories of tests provide different perspectives valuable to understanding the spread of SARS-CoV-2. Within these categories, different biotechnological approaches offer specific advantages and disadvantages. Here we review the categories of tests developed for the detection of the SARS-CoV-2 virus or antibodies against SARS-CoV-2 and discuss the role of diagnostics in the COVID-19 pandemic.

# 2 Importance

Testing is critical to pandemic management. Among molecular tests, messaging about testing strategies has varied widely between countries, with the United States in particular emphasizing the higher sensitivity of polymerase chain reaction tests above immunoassays. However, these tests offer different advantages, and a holistic view of the testing landscape is needed to identify the information provided by each test and its relevance to addressing different questions. Another important consideration is the ease of use and ability to scale for each test, which determines how widely they can be deployed. Here we describe the different diagnostic technologies available as well as the information they provide about SARS-CoV-2 and COVID-19.

# 3 Introduction

Since the emergence of *Severe acute respiratory syndrome-like coronavirus 2* (SARS-CoV-2) in late 2019, significant international efforts have focused on managing the spread of the virus. Identifying individuals who have contracted coronavirus disease 2019 (COVID-19) and may be contagious is crucial to reducing the spread of the virus. Given the high transmissibility of SARS-CoV-2 and the potential for asymptomatic or presymptomatic individuals to be contagious [1], the development of rapid, reliable, and affordable methods to detect SARS-CoV-2 infection is and was vitally important for understanding and controlling spread. For instance, test-trace-isolate procedures were an early cornerstone of many nations' efforts to control the outbreak [2,3,4]. Such efforts depend on diagnostic testing.

The genetic sequence of the SARS-CoV-2 virus was first released by Chinese officials on January 10, 2020 [5], and the first test to detect the virus was released about 13 days later [6]. The genomic information was critical to the development of diagnostic approaches. There are two main classes of diagnostic tests: molecular tests, which can diagnose an active infection by identifying the presence of SARS-CoV-2, and serological tests, which can assess whether an individual was infected in the past via the presence or absence of antibodies against SARS-CoV-2. Over the course of the COVID-19 pandemic, a variety of tests have emerged within these two categories.

Molecular tests detect either viral RNA or protein in a patient sample. They are essential to identifying infected individuals, which can be important for determining courses of action related to treatment, quarantine, and contact tracing. Tests for viral RNA are done by reverse transcription (RT) of viral RNA to DNA followed by DNA amplification, usually with polymerase chain reaction (PCR) [7]. Tests for viral proteins typically use an antibody pair for detection as implemented in techniques such as lateral flow tests (LFTs) and enzyme-linked immunosorbent assays (ELISAs) [8,9]. Molecular tests require the viral genome sequence in order to develop DNA primers for viral RNA detection or to express a viral protein for use as an antigen in antibody production.

Serological tests, on the other hand, detect the presence of antibodies in blood plasma samples or other biological samples, providing insight into whether an individual has acquired immunity against SARS-CoV-2. Assays that can detect the presence of antibodies in blood plasma samples include ELISA, lateral flow immunoassay, and chemiluminescence immunoassay (CLIA) [10]. To distinguish past infection from vaccination, serological tests detect antibodies that bind the nucleocapsid protein of the SARS-CoV-2 virus [11]. They are useful for collecting population-level information for epidemiological analysis, as they can be used to estimate the extent of the infection in a given area. Thus, serological tests may be useful to address population-level questions, such as the percent of cases that manifest as severe versus mild and for guiding public health and economic decisions regarding resource allocation and counter-disease measures.

Molecular and serological tests therefore offer distinct, complementary perspectives on COVID-19 infections. Some of the same technologies are useful to both strategies, and different technologies have been employed to varying extents throughout the world since the start of the COVID-19 pandemic. Two of the primary metrics used to evaluate these tests are sensitivity and specificity. Sensitivity refers to a test's ability to correctly identify a true positive; for example, a test with 50% sensitivity would identify SARS-CoV-2 in only one of every two positive samples. On the other hand, specificity refers to how well a test is able to identify a negative sample as negative. This metric can be relevant both in terms of understanding the risk of false positives and in discussing whether a test is susceptible to identifying other coronaviruses. Here, we review the different types of tests within each category that have been developed and provide perspective on their applications.

# 4     Molecular Tests to Identify SARS-CoV-2

Molecular tests are used to identify distinct genomic subsequences of a viral molecule in a sample or the presence of viral protein, and they thus can be used to diagnose an active viral infection. An important first step is identifying which biospecimens are likely to contain the virus in infected individuals and then acquiring these samples from the patient(s) to be tested.

Common sampling sources for molecular tests include nasopharyngeal cavity samples, such as throat washes, throat swabs, and saliva [12], and stool samples [13]. Once a sample is acquired from a patient, molecular tests detect SARS-CoV-2 based on the presence of either viral nucleic acids or viral proteins.

## 4.1 PCR-Based Tests

When testing for RNA from viruses like SARS-CoV-2, the first step involves pre-processing in order to create complementary DNA (cDNA) from the RNA sample using RT. The second step involves the amplification of a region of interest in the cDNA using successive cycles of heating and cooling. Depending on the application, this amplification is achieved using variations of PCR. Reverse transcription polymerase chain reaction (RT-PCR) tests determine whether a target is present by amplifying a region of interest of cDNA [14]. Some tests use the results of the PCR itself (e.g., a band on a gel) to determine whether the pathogen is present. However, this approach has not been employed widely in diagnostic testing, and instead most PCR-based tests are quantitative.

### 4.1.1 Quantitative Real-Time PCR

In contrast to RT-PCR, quantitative, real-time PCR uses fluorescent dyes that bind to the amplified DNA, thereby allowing a real time assessment of the amplification procedure [14] (in this manuscript we refer to quantitative real-time PCR as qPCR, following the Minimum Information for Publication of Quantitative Real-Time PCR Experiments guidelines [15], and when combined with reverse transcriptase steps, as is required for the evaluation of RNA, it is known as RT-qPCR.) The time resolution provided by qPCR and RT-qPCR is useful because the amount of fluorescence emitted by the sample is proportional to the amount of DNA amplified, and therefore the amount of virus present can be indirectly measured using the cycle threshold ($C_t$) determined by qPCR.

The first test developed and validated for the detection of SARS-CoV-2 used RT-qPCR to detect several regions of the viral genome: the *ORF1b* of the RNA-dependent RNA polymerase (RdRP), the envelope protein gene (*E*), and the nucleocapsid protein gene (*N*) [6]. The publication reporting this test was released on January 23, 2020, less than two weeks after the sequence of the virus was first reported [6]. In collaboration with several other labs in Europe and in China, the researchers confirmed the specificity of this test with respect to other coronaviruses against specimens from 297 patients infected with a broad range of respiratory agents. Specifically, this test uses two probes against RdRP, one of which is specific to SARS-CoV-2 [6]. Importantly, this assay was not found to return false positive results.

In January 2020, Chinese researchers developed a test that used RT-qPCR to identify two gene regions of the viral genome, *ORF1b* and *N* [16]. This assay was tested on samples from two COVID-19 patients and a panel of positive and negative controls consisting of RNA extracted from several cultured viruses. The assay uses the *N* gene to screen patients, while the *ORF1b* gene region is used to confirm the infection [16]. The test was designed to detect sequences conserved across sarbecoviruses, or viruses within the same subgenus as SARS-CoV-2. Considering that *Severe acute respiratory syndrome-related coronavirus 1* (SARS-CoV-1) and

SARS-CoV-2 are the only sarbecoviruses currently known to infect humans, a positive test can be assumed to indicate that the patient is infected with SARS-CoV-2, although this test is not able to discriminate the genetics of viruses within the sarbecovirus clade. The fact that the targets are so conserved offers the advantage of reduced concern about sensitivity in light of the evolution of SARS-CoV-2.

qPCR tests have played an important role in diagnostics during the COVID-19 pandemic. For SARS-CoV-2, studies have typically considered a patient to be infectious if the $C_t$ is below 33 or sometimes 35 [1,17,18]. A lower $C_t$ corresponds to fewer qPCR cycles needed to reach a detectable level, indicating that higher amounts of virus were present in the initial reaction. Interpretations of the $C_t$ values obtained from these tests have raised some interesting questions related to viral load and contagiousness. Lower $C_t$ values correspond to a higher probability of a positive viral culture, but no threshold could discriminate all positive from all negative cultures [19]. Additionally, because of the variability introduced by sample collection and clinical components of testing, $C_t$ is not a proxy for viral load [20]. Positive PCR results have also been reported for extended periods of time from symptom onset and/or the first positive PCR test [21], meaning that in some cases, a positive PCR may not indicate that someone is contagious [1].

In addition to the nuance required to interpret PCR results, there are also factors that influence their accuracy. The specificity of these tests is very high [22], meaning that a positive RT-PCR result is very likely to indicate SARS-CoV-2 infection. The weight given to these tests as an indicator of SARS-CoV-2 infection regardless of other clinical considerations is not typical [23]. In fact, while the analytical specificity of the assay is extremely high, the challenges of implementing testing can introduce variability that results in a lower clinical specificity [23]. Several factors may influence the sensitivity and specificity, with sample collection being a critically important factor in the reliability of RT-PCR results. The most reliable results were found to come from nasopharyngeal swabs and from pooled nasal and throat swabs, with lower accuracies produced by saliva or by throat or nasal swabs alone [22,24]. Differences in experimental parameters such as the use of primers more specific to SARS-CoV-2 has been found to improve sensitivity in these specimens [25]. Additionally, the impact of viral evolution on RT-PCR sensitivity is a concern [26,27]. Using a panel that includes multiple targets can mitigate these effects [28]. Additionally, a test designed to incorporate genomic differences with SARS-CoV-1 was found to offer improved sensitivity and specificity [25]. Thus, while various factors can influence the exact parameters of testing accuracy, RT-PCR is known to have very high specificity and lower, but still high, sensitivity.

### 4.1.2 Digital PCR

Digital PCR (dPCR) is a new generation of PCR technologies offering an alternative to traditional qPCR [29]. In dPCR, a sample is partitioned into thousands of compartments, such as nanodroplets (droplet dPCR or ddPCR) or nanowells, and a PCR reaction takes place in each compartment. This design allows for a digital read-out where each partition is either positive or negative for the nucleic acid sequence being tested for, allowing for absolute target quantification through Poisson statistics. While dPCR equipment is not yet as common as that for qPCR, dPCR for DNA targets generally achieves higher sensitivity than other PCR technologies while maintaining high specificity, though sensitivity is slightly lower for RNA targets [30].

High sensitivity is particularly relevant for SARS-CoV-2 detection, since low viral load in clinical samples can lead to false negatives. In one study, Suo et al. [31] performed a double-blind evaluation of ddPCR for SARS-CoV-2 detection. They evaluated on 63 samples collected from suspected positive outpatients and 14 from supposed convalescent patients. Of the 63 outpatients, only 21 (33%) were identified as positive for SARS-CoV-2 with qPCR. However, ddPCR identified 49 (78%) as positive, 10 (16%) as negative, and 4 (6%) as suspected/borderline for SARS-CoV-2 infection. While both qPCR and ddPCR were found to have very high specificity (100%), this analysis reported that the sensitivity was 40% with qPCR compared to 94% with ddPCR. Analysis of serial dilutions of a linear DNA standard suggested that ddPCR was approximately 500 times more sensitive than qPCR [31]. Thus, this study suggests that ddPCR provides an extremely sensitive molecular test that is able to detect SARS-CoV-2 even at very low viral loads.

A second study [32] confirmed that RT-ddPCR is able to detect SARS-CoV-2 at a lower threshold for viral load relative to RT-PCR. This study analyzed 196 samples, including 103 samples from suspected patients, 77 from contacts and close contacts, and 16 from suspected convalescents, using both RT-qPCR and RT-ddPCR. First, the authors evaluated samples from the 103 suspected cases. Using RT-qPCR, 29 (28%) were identified as positive, 25 (24%) as negative, and 49 (48%) as borderline, i.e., the $C_t$ value was higher than the positive threshold of 35 but lower than the negative threshold of 40. When the 61 negative and borderline samples were reanalyzed with ddPCR, 19 (31%) of the negative and 42 (69%) of the borderline samples were identified as positive. All of the suspected cases were later confirmed to be COVID-19 through a combination of symptom development and RT-qPCR resampling, indicating that ddPCR improved the overall detection rate compared to RT-qPCR from 28.2% to 87.4%.

They repeated this analysis in patient samples from contacts and close contacts. Patients who tested negative with both methods (n = 48) were observed to remain healthy over a period of 14 days. Among the remaining 29 samples from contacts, RT-qPCR identified 12 as positive, 1 as negative, and 16 as borderline. All of the samples that tested positive using RT-qPCR also tested positive using ddPCR. In contrast, the negative result and all but one of the borderline results were identified as positive by RT-ddPCR, and these patients were later determined to be SARS-CoV-2 positive based on clinical evaluation and repeated molecular sampling. Similarly, in the final group, 16 convalescent patients, RT-qPCR identified 12 as positive, three as suspect, and one as negative, but RT-dPCR identified all as positive. The evidence from this study therefore supports a lower limit of detection with ddPCR. Overall, these studies suggest that ddPCR is a promising tool for overcoming the problem of false negatives in SARS-CoV-2 RNA testing, but this method is unlikely to affect the current pandemic due to its lack of availability.

### 4.1.3 Sequencing

In some cases, the DNA amplified with PCR is sequenced. Sequencing requires an additional sample pre-processing step called library preparation. Library preparation is the process of preparing the sample for sequencing, typically by fragmenting the sequences and adding adapters [33]. In some cases, library preparation can involve other modifications of the sample, such as adding barcodes to identify a particular sample within the sequence data. Barcoding can therefore be used to pool samples from multiple sources. There are different reagents used for library preparation that are specific to identifying one or more target sections with PCR [34]. Sequential pattern matching is then used to identify unique subsequences of the virus, and if sufficient subsequences are found, the test is considered positive. Therefore, tests that use sequencing require a number of additional molecular and analytical steps relative to tests that use PCR alone.

Sequencing has been an important strategy for discovery of SARS-CoV-2 variants (e.g., see [35]). Sequencing elucidates any genetic variants located between the PCR primers. For this reason, it is critical to genomic surveillance efforts. Genomic surveillance is an important complement to epidemiological surveillance efforts [36], as described below. Through genomic surveillance, it has become possible to monitor the emergence of variants of interest and variants of concern (VOC) that may pose additional threats due to increased contagiousness, virulence, or immune escape [36,37]. Sequencing also allows for analysis of the dominant strains in an area at a given time. Worldwide, the extent of genomic surveillance varies widely, with higher-income countries typically able to sequence a higher percentage of cases [38]. Sequencing efforts are important for identifying variants containing mutations that might affect the reliability of molecular diagnostic tests, as well as mitigation measures such as therapeutics and prophylactics [26,27]. Therefore, sequencing is an important component of diagnostics: while it is not necessary for diagnosing an individual case, it is critical to monitoring trends in the variants affecting a population and to staying aware of emerging variants that may pose additional challenges.

### 4.1.4 Pooled and Automated PCR Testing

Due to limited supplies and the need for more tests, several labs have found ways to pool or otherwise strategically design tests to increase throughput. The first such result came from Yelin et al. [39], who reported that they could pool up to 32 samples in a single qPCR run. This was followed by larger-scale pooling with slightly different methods [40]. Although these approaches are also PCR based, they allow for more rapid scaling and higher efficiency for testing than the initial PCR-based methods developed. Conceptually, pooling could also be employed in analysis with RT-qPCR [41], and this strategy has been evaluated in settings such as schools [42] and hospitals [43].

## 0.4.2 RT-LAMP

RT-PCR remains the gold standard for detection of SARS-CoV-2 RNA from infected patients, but the traditional method requires special equipment and reagents, including a thermocycler. Loop-mediated isothermal amplification (LAMP) is an alternative to PCR that does not require specialized equipment [44]. In this method, nucleic acids are amplified in a 25 µL reaction that is incubated and chilled on ice [44]. It uses primers designed to facilitate auto-cycling strand displacement DNA synthesis [44]. LAMP can be combined with reverse transcription (RT-LAMP) to enable the detection of RNA.

One study showed that RT-LAMP is effective for detection of SARS-CoV-2 with excellent specificity and sensitivity and that this method can be applied to unprocessed saliva samples [45]. This method was benchmarked against RT-PCR using 177 human nasopharyngeal RNA samples, of which 126 were COVID positive. The authors break down the sensitivity of their test according to the $C_t$ value from RT-PCR of the same samples; RT-LAMP performs at 100% sensitivity for samples with a $C_t$ from RT-PCR of 32 or less. The performance is worse when considering all RT-PCR positive samples (including those with $C_t$ values between 32-40). However, there is some evidence suggesting that samples obtained from individuals that achieve $C_t$ values >30 measured using RT-PCR tend to be less infective that those that record a $C_t$ value <30 [46,47,48], so RT-LAMP is still a useful diagnostic tool. Various combinations of reagents are available, but one example is the WarmStart Colorimetric LAMP 2X Master Mix with a set of six primers developed previously by Zhang et al. [49]. To determine assay sensitivity, serial tenfold dilutions of *in vitro* transcribed *N*-gene RNA standard were tested using quantities from $10^5$ copies down to 10 copies. The assay readout is the color of the dye changing from pink to yellow due to binding to the DNA product over 30 minutes. The RT-LAMP assay was then applied to clinical nasopharyngeal samples. For viral loads above 100 copies of genomic RNA, the RT-LAMP assay had a sensitivity of 100% and a specificity of 96.1% from purified RNA. The sensitivity of the direct assay of saliva by RT-LAMP was 85%. Sensitivity and specificity metrics were obtained by comparison with results from RT-PCR. RT-LAMP pilot studies for detection of SARS-CoV-2 were reviewed in a meta-analysis [50]. In the meta-analysis of all 2,112 samples, the cumulative sensitivity of RT-LAMP was calculated at 95.5%, and the cumulative specificity was 99.5%.

This test aims to bring the sensitivity of nucleic acid detection to the point of care or home testing setting. It could be applied for screening, diagnostics, or as a definitive test for people who are positive based on LFTs (see below). The estimated cost per test is about 2 euros when RNA extraction is included. The main strength of this test over RT-PCR is that it can be done isothermally, but the main drawback is that it is about 10-fold less sensitive than RT-PCR. The low cost, excellent sensitivity/specificity, and quick readout of RT-LAMP makes this an attractive alternative to RT-PCR. Alternative strategies like RT-LAMP are needed to bring widespread testing away from the lab and into under-resourced areas.

## 4.3   CRISPR-based Detection

Technology based on CRISPR (clustered regularly interspaced short palindromic repeats) [51] has also been instrumental in scaling up testing protocols. Two CRISPR-associated nucleases, Cas12 and Cas13, have been used for nucleic acid detection. Multiple assays exploiting these nucleases have emerged as potential diagnostic tools for the rapid detection of SARS-CoV-2 genetic material and therefore SARS-CoV-2 infection. The SHERLOCK method (Specific High-sensitivity Enzymatic Reporter unLOCKing) from Sherlock Biosciences relies on Cas13a to discriminate between inputs that differ by a single nucleotide at very low concentrations [52]. The target RNA is amplified by real-time recombinase polymerase amplification (RT-RPA) and T7 transcription, and the amplified product activates Cas13a. The nuclease then cleaves a reporter RNA, which liberates a fluorescent dye from a quencher. Several groups have used the SHERLOCK method to detect SARS-CoV-2 viral RNA. An early study reported that the method could detect 7.5 copies of viral RNA in all 10 replicates, 2.5 copies in 6 out of 10, and 1.25 copies in 2 out of 10 runs [53]. It also reported 100% specificity and sensitivity on 114 RNA samples from clinical respiratory samples (61 suspected cases, among which 52 were confirmed and nine were ruled out by metagenomic next-generation sequencing, 17 SARS-CoV-2-negative but human coronavirus (HCoV)-positive cases, and 36 samples from healthy subjects) and a reaction turnaround time of 40 minutes. A separate study screened four designs of SHERLOCK and extensively tested the best-performing assay. They determined the limit of detection to be 10 copies/µl using both fluorescent and lateral flow detection [54].

LFT strips are simple to use and read, but there are limitations in terms of availability and cost per test. Another group therefore proposed the CREST (Cas13-based, Rugged, Equitable, Scalable Testing) protocol, which uses a P51 cardboard fluorescence visualizer, powered by a 9-volt battery, for the detection of Cas13 activity instead of immunochromatography [55]. CREST can be run, from RNA sample to result, with no need for AC power or a dedicated facility, with minimal handling in approximately 2 hours. Testing was performed on 14 nasopharyngeal swabs. CREST picked up the same positives as the CDC-recommended TaqMan assay with the exception of one borderline sample that displayed low-quality RNA. This approach may therefore represent a rapid, accurate, and affordable procedure for detecting SARS-CoV-2.

The DETECTR (DNA Endonuclease-Targeted CRISPR Trans Reporter) method from Mammoth Biosciences involves purification of RNA extracted from patient specimens, amplification of extracted RNAs by loop-mediated amplification, and application of their Cas12-based technology. In this assay, guide RNAs (gRNAs) were designed to recognize portions of sequences corresponding to the SARS-CoV-2 genome, specifically the N2 and E regions [56]. In the presence of SARS-CoV-2 genetic material, sequence recognition by the gRNAs results in double-stranded DNA cleavage by Cas12, as well as cleavage of a single-stranded DNA molecular beacon. The cleavage of this molecular beacon acts as a colorimetric reporter that is subsequently read out in a lateral flow assay and indicates the presence of SARS-CoV-2 genetic material and therefore SARS-CoV-2 infection. The 40-minute assay is considered positive if there is detection of both the *E* and *N* genes or presumptive positive if there is detection of either of them. The assay had 95% positive predictive agreement and 100% negative predictive agreement with the US Centers for Disease Control and Prevention SARS-CoV-2 RT-qPCR assay. The estimated limit of detection was 10 copies per µl reaction, versus 1 copy per µl reaction for the CDC assay.

These results have been confirmed by other DETECTR approaches. Using RT-RPA for amplification, another group detected 10 copies of synthetic SARS-CoV-2 RNA per µl of input within 60 minutes of RNA sample preparation in a proof-of-principle evaluation [57]. Through a similar approach, another group reported detection at 1 copy per µl [58]. The DETECTR protocol was improved by combining RT-RPA and CRISPR-based detection in a one-pot reaction that incubates at a single temperature and by using dual CRISPR RNAs, which increases sensitivity. This new assay, known as All-In-One Dual CRISPR-Cas12a, detected 4.6 copies of SARS-CoV-2 RNA per µl of input in 40 minutes [59]. Another single-tube, constant-temperature approach using Cas12b instead of Cas12a achieved a detection limit of 5 copies/µl in 40-60 minutes [60].

It was also reported that electric field gradients can be used to control and accelerate CRISPR assays by co-focusing Cas12-gRNA, reporters, and target [61]. The authors generated an appropriate electric field gradient using a selective ionic focusing technique known as isotachophoresis (ITP) implemented on a microfluidic chip. They also used ITP for automated purification of target RNA from raw nasopharyngeal swab samples. Combining this ITP purification with loop-mediated isothermal amplification, their ITP-enhanced assay achieved detection of SARS-CoV-2 RNA (from raw sample to result) in 30 minutes.

All these methods require upstream nucleic acid amplification prior to CRISPR-based detection. They rely on type V (Cas12-based) and type IV (Cas13-based) CRISPR systems. In contrast, type III CRISPR systems have the unique property of initiating a signaling cascade, which could boost the sensitivity of direct RNA detection. In type III CRISPR systems, guide CRISPR RNAs (crRNAs) are bound by several Cas proteins [62] and can target both DNA and RNA molecules [63,64]. A study tested this hypothesis using the type III-A crRNA-guided surveillance complex from *Thermus thermophilus* [65]. The authors showed that activation of the Cas10 polymerase generates three products (cyclic nucleotides, protons, and pyrophosphates) that can all be used to detect SARS-CoV-2 RNA. Detection of viral RNA in patient samples still required an initial nucleic acid amplification step, but improvements may in the future remove that requirement.

This goal of amplification-free detection was later achieved for a Cas13a-based system [66]. This approach combined multiple CRISPR RNAs to increase Cas13a activation, which is detected by a fluorescent reporter. Importantly, because the viral RNA is detected directly, the test yields a quantitative measurement rather than a binary result. The study also shows that fluorescence can be measured in a custom-made dark box with a mobile phone camera and a low-cost laser illumination and collection optics. This approach is a truly portable assay for point-of-care diagnostics. The authors achieved detection of 100 copies/µl of pre-isolated RNA in 30 minutes, and correctly identified all SARS-CoV-2-positive patient RNA samples tested in 5 minutes (n = 20).

There is an increasing body of evidence that CRISPR-based assays offer a practical solution for rapid, low-barrier testing in areas that are at greater risk of infection, such as airports and local community hospitals. In the largest study to date, DETECTR was compared to RT-qPCR on 378 patient samples [67]. The authors reported 95% reproducibility. Both techniques were equally sensitive in detecting SARS-CoV-2. Lateral flow strips showed 100% correlation to the high-throughput DETECTR assay. Importantly, DETECTR was 100% specific for SARS-CoV-2 and did not detect other human coronaviruses. A method based on a Cas9 ortholog from *Francisella novicida* known as FnCas9 achieved 100% sensitivity and 97% specificity in clinical samples, and the diagnostic kit is reported to have completed regulatory validation in India [68].

## 4.4   Immunoassays for the Detection of Antigens

Immunoassays can detect molecular indicators of SARS-CoV-2 infection, such as the proteins that act as antigens from the SARS-CoV-2 virus. They offer the advantage of generally being faster and requiring less specialized equipment than other molecular tests, especially those involving PCR. As a result, immunoassays hold particular interest for implementation at home and in situations where resources for PCR testing are limited. The trade-off is that these tests typically have a lower sensitivity, and sometimes a lower specificity, than other molecular tests. However, these tests tend to return a positive result five to 12 days after symptom onset, which may therefore correlate more closely with the timeframe during which viral replication occurs [69]. Immunoassays for the detection of the SARS-CoV-2 antigen can include LFTs and ELISA, as discussed here, as well as CLIA and chromatographic immunoassays [70], as described in the serological testing section below.

### 4.4.1 Lateral Flow Tests

LFTs provide distinct value relative to PCR tests. They can return results within 30 minutes and can be performed without specialized equipment and at low cost. They also do not require training to operate and are cheap to produce. Thus, they can be distributed widely to affected populations making them an important public health measure to curb pandemic spread. LFTs rely on the detection of viral protein with an antibody. Often this is done with an antibody sandwich format, where one antibody conjugated to a dye binds at one site on the antigen, and an immobilized antibody on the strip binds at another site [8]. This design allows the dye to accumulate to form a characteristic positive test line on the strip [8]. Outside of COVID-19 diagnostics, the applications of LFTs are broad; they are routinely used for home pregnancy tests, disease detection, and even drugs of abuse detection in urine [71].

A recent review surveyed the performance of LFTs for detection of current SARS-CoV-2 infection [72]. This review covered 24 studies that included more than 26,000 total LFTs. They reported significant heterogeneity in test sensitivities, with estimates ranging from 37.7% to 99.2%. The estimated specificities of these tests were more homogeneous, spanning 92.4% to 100.0%.

Despite having lower sensitivity than PCR tests, LFTs occupy an important niche in the management of SARS-CoV-2. Current infection detection by LFTs enables the scale and speed of testing that is beneficial to managing viral spread. LFTs were available freely to citizens in the United Kingdom until April 1, 2022 [73] and to citizens of the United States in early 2022 [74]. These tests are particularly useful for ruling out SARS-CoV-2 infection in cases where the likelihood of infection is low (e.g., asymptomatic individuals) and positives (including false positives) can be validated with testing by alternate means [75].

### 4.4.2 Enzyme-Linked Immunosorbent Assay

ELISA is a very sensitive immunoassay that can be considered a gold standard for the detection of biological targets, including antibodies and antigenic proteins [9]. It can be used to generate either quantitative or qualitative results that can be returned within a few hours [76]. ELISA builds on the idea that antibodies and antigens bind together to form complexes [9] and utilizes an enzyme covalently linked to an antibody against the antigen to produce assay signal, usually a color change [77]. The main advantage of ELISA is that it enables signal amplification through the enzyme's activity, which increases sensitivity. With sandwich ELISA, antibodies are immobilized on a surface such as a plate, and viral protein antigens in the sample bind and are retained [78]. A second antibody is added that binds to another site on the antigen is then added, and that second antibody is covalently linked to an enzyme. A substrate for that enzyme is then added to produce signal, usually light or a color change The exact strategy for tagging with a reporter enzyme varies among different types of ELISA [9,78]. For COVID-19 diagnostics, ELISAs have been designed to detect the antigenic Spike protein [79].

One of these assays uses two monoclonal antibodies specific to the nucleocapsid of SARS-CoV-2 to evaluate the relationship between the effect of (estimated) viral load on the ability of the assay to detect the SARS-CoV-2 antigen [80]. This study analyzed 339 naso-oropharyngeal samples that were also analyzed with RT-qPCR as a gold standard. RT-qPCR identified 147 samples as positive and 192 as negative. The authors estimated the overall sensitivity and specificity to be 61.9% and 99.0%, respectively. Sensitivity increased with higher $C_t$. This study also assessed the performance of the ELISA test under different conditions in order to evaluate how robust it would be to the challenges of testing in real-world settings globally. Higher sensitivity was achieved for samples that were stored under ideal conditions (immediate placement in -80° C). Therefore, while immediate access to laboratory equipment is an advantage, it is not strictly necessary for ELISA to detect the antigen.

## 4.5 Limitations of Molecular Tests

Tests that identify SARS-CoV-2 using molecular technologies will identify only individuals with current infections and are not appropriate for identifying individuals who have recovered from a previous infection. Among molecular tests, different technologies have different sensitivities and specificities. In general, specificity is high, and even then, the public health repercussions of a false positive can generally be mitigated with follow-up testing. On the other hand, a test's sensitivity, which indicates the risk of a false-negative response, can pose significant challenge to large-scale testing. False negatives are a significant concern for several reasons. Importantly, clinical reports indicate that it is imperative to exercise caution when interpreting the results of molecular tests for SARS-CoV-2 because negative results do not necessarily mean a patient is virus-free [81]. To reduce occurrence of false negatives, correct execution of the analysis is crucial [82]. Additionally, PCR-based tests can remain positive for a much longer time than the virus is likely to be actively replicating [69], raising concerns about their informativeness after the acute phase of the disease. Hence, the CDC has advised individuals who suspect they have been re-infected with SARS-CoV-2 to avoid using diagnostic tests within 90 days of receiving a previous positive test [83].

Additionally, the emerging nature of the COVID-19 pandemic has introduced some challenges related to uncertainty surrounding interactions between SARS-CoV-2 and its human hosts. For example, viral shedding kinetics are still not well understood but are expected to introduce a significant effect of timing of sample collection on test results [82]. Similarly, the type of specimen could also influence outcomes, as success in viral detection varies among clinical sample types [22,24,82]. With CRISPR-based testing strategies, the gRNA can recognize off-target interspersed sequences in the viral genome [84], potentially resulting in false positives and a loss of specificity.

There are also significant practical and logistical concerns related to the widespread deployment of molecular tests. Much of the technology used for molecular tests is expensive, and while it might be available in major hospitals and/or diagnostic centers, it is often not available to smaller facilities [85]. At times during the pandemic, the availability of supplies for testing, including swabs and testing media, has also been limited [86]. Similarly, processing times can be long, and tests might take up to 4 days to return results [85], especially during times of high demand, such as spikes in case numbers [87]. Countries have employed various and differing molecular testing strategies as a tool to reduce viral transmission, even among high-income countries [88]. The rapid development of molecular tests has provided a valuable, albeit imperfect, tool to identify active SARS-CoV-2 infections.

# 5   Serological Tests to Identify Recovered Individuals

Although several molecular diagnostic tests to detect viral genetic material have high specificity and sensitivity, they provide information only about active infection, and therefore offer just a snapshot-in-time perspective on the spread of a disease. Most importantly, they would not work on a patient who has fully recovered from the virus at the time of sample collection. In such contexts, serological tests are informative.

Serological tests use many of the same technologies as the immunoassays used to detect the presence of an antigen but are instead used to evaluate the presence of antibodies against SARS-CoV-2 in a serum sample. These tests are particularly useful for insight into population-level dynamics and can also offer a glimpse into the development of antibodies by individual patients during the course of a disease. Immunoassays can detect antibodies produced by the adaptive immune system in response to viral threat. Understanding the acquisition and retention of antibodies is important both to the diagnosis of prior (inactive) infections and to the development of vaccines. The two immunoglobulin classes that are most pertinent to these goals are immunoglobulin M (IgM), which are the first antibodies produced in response to an infection, and immunoglobulin G (IgG), which are the most abundant antibodies [89,90]. Serological tests detect these antibodies, offering a mechanism through which prior infection can be identified. However, the complexity of the human immune response means that there are many facets to such analyses.

In general, SARS-CoV-2 infection will induce the immune system to produce antibodies fairly quickly. Prior research is available about the development of antibodies to SARS-CoV-1 during the course of the associated disease, severe acute respiratory syndrome (SARS). IgM and IgG antibodies were detected in the second week following SARS-CoV-1 infection. IgM titers peaked by the first month post-infection, and then declined to undetectable levels after day 180. IgG titers peaked by day 60 and persisted in all donors through the two-year duration of study [91]. Such tests can also illuminate the progression of viral disease, as IgM are the first antibodies produced by the body and indicate that the infection is active. Once the body has responded to the infection, IgG are produced and gradually replace IgM, indicating that the body has developed immunogenic memory [92]. Therefore, it was hoped that the development of assays to detect the presence of IgM and IgG antibodies against SARS-CoV-2 would allow the identification of cases from early in the infection course (via IgM) and for months or years afterwards (via IgG). Several technologies have been used to develop serological tests for COVID-19, including ELISA, lateral flow immunoassay, chemiluminescence immunoassay, and neutralizing antibody assays [93].

## 5.1 ELISA

The application of ELISA to serological testing is complementary to its use in molecular diagnostics (see above). Instead of using an enzyme-labeled antibody as a probe that binds to the target antigen, the probe is an antigen and the target is an antibody. The enzyme used for detection and signal amplification is on a secondary antibody raised generally against human IgG or IgM. In March 2020, the Krammer lab proposed an ELISA test that detects IgG and IgM that react against the receptor-binding domain (RBD) of the spike proteins (S) of the virus [94]. A subsequent ELISA test developed to detect SARS-CoV-2 IgG based on the RBD reported a specificity of over 99% and a sensitivity of up to 88.24%, which was observed in samples collected 21 to 27 days after the onset of infection (approximated with symptom onset or positive PCR test) [95]. Earlier in the disease course, sensitivity was lower: 53.33% between days 0 and 13 and 80.47% between days 14 and 20. This study reported that their laboratory ELISA outperformed two commercial kits that also used an ELISA design [95]. Therefore, while analysis with ELISA requires laboratory support and equipment, these results do suggest that ELISA achieves relatively high sensitivity, especially in the weeks following infection. Efforts have been made to develop low-cost strategies for conducting these tests that will make them more accessible worldwide [96].

## 5.2 Chemiluminescence Immunoassay

Another early approach investigated for detection of antibodies against SARS-CoV-2 was CLIA. Like ELISA, CLIA is a type of enzyme immunoassay (EIA) [97]. While the technique varies somewhat, in one approach, a bead is coated with the antigen and then washed with the sample [98]. If the antibody is present in the sample, it will bind to the bead. Then the bead is exposed to a label, a luminescent molecule that will bind to the antigen/antibody complex and can therefore be used as an indicator [98]. One CLIA approach to identify COVID-19 used a synthetic peptide derived from the amino acid sequence of the SARS-CoV-2 S protein [99]. It was highly specific to SARS-CoV-2 and detected IgM in 57.2% and IgG in 71.4% of serum samples from 276 COVID-19 cases confirmed with RT-qPCR. IgG could be detected within two days of the onset of fever, but IgM could not be detected any earlier [99], which has been supported by other analyses as well [100]. This pattern was consistent with observations in Middle East respiratory syndrome, which is also caused by an HCoV. In comparisons of different commercial immunoassays, accuracy of CLIA tests were often roughly comparable to other EIAs [101], although one CLIA did not perform as well as several other EIAs [100,102]. The sensitivity and specificities reported vary among CLIA tests and for the detection of IgM versus IgG, but sensitivities and specificities as high as 100% have been reported among various high-throughput tests [102,103,104]. CLIA has previously been used to develop tests that can be used at point of care (e.g., [97]) which may allow for this technique to become more widely accessible in the future.

## 5.3 Lateral Flow Immunoassay

The first serological test approved for emergency use in the United States was developed by Cellex [105]. The Cellex qSARS-CoV-2 IgG/IgM Rapid Test is a chromatographic immunoassay, also known as a lateral flow immunoassay, designed to qualitatively detect IgM and IgG antibodies against SARS-CoV-2 in the plasma of patients suspected to have developed a SARS-CoV-2 infection [105]. The Cellex test cassette contains a pad of SARS-CoV-2 antigens and a nitrocellulose strip with lines for each of IgG and IgM, as well as a control (goat IgG) [105]. In a specimen that contains antibodies against the SARS-CoV-2 antigen, the antibodies will bind to the strip and be captured by the IgM and/or IgG line(s), resulting in a change of color [105]. With this particular assay, results can be read within 15 to 20 minutes [105]. Lateral flow immunoassays are often available at point of care but can have very low sensitivity [102].

## 5.4 Neutralizing Antibody Assays

Neutralizing antibody assays play a functional role in understanding immunity that distinguishes them from other serological tests. The tests described above are all binding antibody tests. On the other hand, rather than simply binding an antibody to facilitate detection, neutralizing antibody assays determine whether an antibody response is present that would prevent infection [106,107]. Therefore, these tests serve the purpose of evaluating the extent to which a sample donor has acquired immunity that will reduce susceptibility to SARS-CoV-2. As a result, neutralizing antibody assays have been used widely to characterize the duration of immunity following infection, to assess vaccine candidates, and to establish correlates of protection against infection and disease [108,109,110]. These tests are typically performed in a laboratory [106], and in SARS-CoV-2, the results of neutralizing antibody assays are often correlated with the results of binding antibody tests [106].

The gold standard for assessing the presence of neutralizing antibodies is the plaque reduction neutralization test (PRNT), but this approach does not scale well [107]. An early high-throughput neutralizing antibody assay designed against SARS-CoV-2 used a fluorescently labeled reporter virus that was incubated with different dilutions of patient serum [107]. The cells used for incubation would turn green if antibodies were not present. Essentially, this assay evaluates whether the virus is able to infect the cell in the presence of the serum. The specificity of this assay was 100%, and the correlation between the results of this assay and of PRNT was 0.85 with the results suggesting that the sensitivity of the high-throughput approach was higher than that of PRNT [107]. While this approach was performed on a plate and using cells, other methods have been developed using methods such as bead arrays [111].

## 5.5 Duration of Immune Indicators

While the adaptive immune system produces antibodies in response to SARS-CoV-2 viral challenge, these indicators of seroconversion are unlikely to remain in circulation permanently. Previously, a two-year longitudinal study following convalesced SARS patients with a mean age of 29 found that IgG antibodies were detectable in all 56 patients surveyed for at least 16 months and remained detectable in all but 4 patients (11.8%) through the full two-year study period [112]. These results suggest that immunity to SARS-CoV-1 is sustained for at least a year. Circulating antibody titers to other coronaviruses have been reported to decline significantly after one year [113]. Evidence to date suggests that sustained immunity to the SARS-CoV-2 virus remains for a shorter period of time but at least 6 to 8 months after infection

[114,115,116,117]. However, this does not mean that all serological evidence of infection dissipates, but rather that the immune response becomes insufficient to neutralize the virus.

In order to study the persistence of SARS-CoV-2 antibodies, one study assessed sustained immunity using 254 blood samples from 188 COVID-19 positive patients [115]. The samples were collected at various time points between 6 and 240 days post-symptom onset; some patients were assessed longitudinally. Of the samples, 43 were collected at least 6 months after symptom onset. After one month, 98% of patients were seropositive for IgG to S. Moreover, S IgG titers were stable and heterogeneous among patients over a period of 6 to 8 months post-symptom onset, with 90% of subjects seropositive at 6 months. Similarly, at 6 to 8 months 88% of patients were seropositive for RBD IgG, and 90% were seropositive for SARS-CoV-2 neutralizing antibodies. Another study examined 119 samples from 88 donors who had recovered from mild to severe cases of COVID-19 [117]. A relatively stable level of IgG and plasma neutralizing antibodies was identified up to 6 months post diagnosis. Significantly lower but considerable levels of anti-SARS-CoV-2 IgG antibodies were still present in 80% of samples obtained 6 to 8 months post-symptom onset.

Titers of IgM and IgG antibodies against the RBD were found to decrease from 1.3 to 6.2 months post infection in a study of 87 individuals [118]. However, the decline of IgA activity (15%) was less pronounced than that of IgM (53%) or IgG (32%). It was noted that higher levels of anti-RBD IgG and anti-N total antibodies were detected in individuals that reported persistent post-acute symptoms at both study visits. Moreover, plasma neutralizing activity decreased five-fold between 1.3 and 6.2 months in an assay of HIV-1 virus pseudotyped with SARS-CoV-2 S protein, and this neutralizing activity was directly correlated with IgG anti-RBD titers [118]. These findings are in accordance with other studies that show that the majority of seroconverters have detectable, albeit decreasing, levels of neutralizing antibodies at least 3 to 6 months post infection [119,120,121].

Determining the potency of anti-RBD antibodies early in the course of an infection may be important moving forward, as their neutralizing potency may be prognostic for disease severity and survival [122]. The duration of immunity might also vary with age [123] or ABO blood type [124]. Autopsies of lymph nodes and spleens from severe acute COVID-19 patients showed a loss of T follicular helper cells and germinal centers that may explain some of the impaired development of antibody responses [125]. Therefore, serological testing may be time-limited in its ability to detect prior infection.

Other immune indicators of prior infection have also been evaluated to see how they persist over time. SARS-CoV-2 memory $CD8^+$ T cells were slightly decreased (50%) 6 months post-symptom onset. In this same subset of COVID-19 patients, 93% of subjects had detectable levels of SARS-CoV-2 memory $CD4^+$ T cells, of which 42% had more than 1% SARS-CoV-2-specific $CD4^+$ T cells. At 6 months, 92% of patients were positive for SARS-CoV-2 memory $CD4^+$ T cells. Indeed, the abundance of S-specific memory $CD4^+$ T cells over time was similar to that of SARS-CoV-2-specific $CD4^+$ T cells overall [115]. T cell immunity to SARS-CoV-2 at 6 to 8 months following symptom onset has also been confirmed by other studies [117,126,127]. In another study, T cell reactivity to SARS-CoV-2 epitopes was also detected in some individuals never been exposed to SARS-CoV-2. This finding suggests the potential for cross-reactive T cell recognition between SARS-CoV-2 and pre-existing circulating HCoV that are responsible for the "common cold" [128], but further research is required. Therefore, whether T cells will provide a more stable measure through which to assess prior infection remains unknown. Notably, commercial entities have tried to develop tests specifically for T cells, some of which have been authorized by the United States Food and Drug Administration [129,130] to identify people with adaptive T cell immune responses to SARS-CoV-2, either from a previous or ongoing infection.

## 5.6 Applications of Serological Tests

In addition to the limitations posed by the fact that antibodies are not permanent indicators of prior infection, serological immunoassays carry a number of limitations that influence their utility in different situations. Importantly, false positives can occur due to cross-reactivity with other antibodies according to the clinical condition of the patient [105]. Due to the long incubation times and delayed immune responses of infected patients, serological immunoassays are insufficiently sensitive for a diagnosis in the early stages of an infection. Therefore, such tests must be used in combination with RNA detection tests if intended for diagnostic purposes [131]. False positives are particularly harmful if they are erroneously interpreted to mean that a population is more likely to have acquired immunity to a disease [132]. Similarly, while serological tests may be of interest to individuals who wish to confirm they were infected with SARS-CoV-2 in the past, their potential for false positives means that they are not currently recommended for this use. However, in the wake of vaccines becoming widely available, accurate serological tests that could be administered at point of care were investigated in the hope that they could help to prioritize vaccine recipients [133]. Another concern with serological testing is the potential for viral evolution to reduce the sensitivity of assays, especially for neutralizing antibody assays. Chen et al. performed a systematic re-analysis of published data examining the neutralizing effect of serum from vaccinated or recovered individuals on four VOC [134]. They found reduced neutralizing titers against these variants relative to the lineages used for reference. These findings suggest that such techniques will need to be modified over time as SARS-CoV-2 evolves.

These limitations make serological tests far less useful for diagnostics and for test-and-trace strategies; however, serological testing is valuable for public health monitoring at the population level. Serosurveys provide a high-level perspective of the prevalence of a disease and can provide insight into the susceptibility of a population as well as variation in severity, e.g., between geographic regions [132]. From a public health perspective, they can also provide insight into the effectiveness of mitigation efforts and to gain insight into risk factors influencing susceptibility [135]. EIA methods are high-throughput [136,137], and, as with molecular tests, additional efforts have been made to scale up the throughput of serological tests [138]. Therefore, serological tests can be useful to developing strategies for the management of viral spread.

Early in the course of the pandemic, it was also hoped that serological tests would provide information relevant to advancing economic recovery. Some infectious agents can be controlled through "herd immunity", which is when a critical mass within the population acquires immunity through vaccination and/or infection, preventing an infectious agent from spreading widely. It was hoped that people who had recovered and developed antibodies might be able to return to work [139,140]. This strategy would have relied on recovered individuals acquiring long-term immunity, which has not been borne out [141]. Additionally, it was hoped that identifying seroconverters and specifically those who had mounted a strong immune response would reveal strong candidates for convalescent plasma donation [94]; however, convalescent plasma has not been found to offer therapeutic benefit (reviewed in [142]). While these hopes have not been realized, serological tests have been useful for gaining a better understanding of the pandemic [135].

# 6  Possible Alternatives to Current Diagnostic Practices

One possible alternative or complement to molecular and serological testing would be diagnosing COVID-19 cases based on symptomatology. COVID-19 can present with symptoms similar to other types of pneumonia, and symptoms can vary widely among COVID-19 patients; therefore, clinical presentation is often insufficient as a sole diagnostic criterion. In addition, identifying and isolating mild or asymptomatic cases is critical to efforts to manage outbreaks. Even among mildly symptomatic patients, a predictive model based on clinical symptoms had a sensitivity of only 56% and a specificity of 91% [143]. More problematic is that clinical symptom-based tests are only able to identify already symptomatic cases, not presymptomatic or asymptomatic cases. They may still be important for clinical practice and for reducing tests needed for patients deemed unlikely to have COVID-19.

In some cases, clinical signs may also provide information that can inform diagnosis. Using computed tomography of the chest in addition to RT-qPCR testing was found to provide a higher sensitivity than either measure alone [144]. X-ray diagnostics have been reported to have high sensitivity but low specificity in some studies [145]. Other studies have shown that specificity varies between radiologists [146], though the sensitivity reported here was lower than that published in the previous paper. While preliminary machine-learning results suggested that chest X-rays might provide high sensitivity and specificity and potentially facilitate the detection of asymptomatic and presymptomatic infections (e.g., [147]), further investigation suggested that such approaches are prone to bias and are unlikely to be clinically useful [148]. Given the above, the widespread use of X-ray tests on otherwise healthy adults is likely inadvisable.

Finally, in addition to genomic and serological surveillance, other types of monitoring have proven useful in managing the pandemic [149]. One that has received significant attention is wastewater surveillance. This approach can use several of the technologies described for molecular testing, such as qPCR and dPCR, as well as *in vitro* culturing [150] and can provide insight into trends in the prevalence of SARS-CoV-2 regionally.

# 7  Strategies and Considerations for Testing

Deciding whom to test, when to test, and which test to use have proven challenging as the COVID-19 pandemic has unfolded. Early in the COVID-19 pandemic, testing was typically limited to individuals considered high risk for developing serious illness [151]. This approach often limited testing to people with severe symptoms and people showing mild symptoms that had been in contact with a person who had tested positive. Individuals who were asymptomatic

(i.e., potential spreaders) and individuals who were able to recover at home were thus often unaware of their status. Therefore, this method of testing administration misses a high proportion of infections and does not allow for test-and-trace methods to be used. For instance, a study from Imperial College estimates that in Italy, the true number of infections was around 5.9 million in a total population of ~60 million, compared to the 70,000 detected as of March 28, 2020 [152]. Another analysis, which examined New York state, indicated that as of May 2020, approximately 300,000 cases had been reported in a total population of approximately 20 million [153]. This corresponded to ~1.5% of the population, but ~12% of individuals sampled statewide were estimated as positive through antibody tests (along with indications of spatial heterogeneity at higher resolution) [153]. Technological advancements that facilitate widespread, rapid testing would therefore be valuable for accurately assessing the rate of infection and aid in controlling the virus' spread. Additionally, the trade off of accessibility, sensitivity, and time to results has raised some complex questions around which tests are best suited to certain situations. Immunoassays, including serological tests, have much higher limits of detection than PCR tests do [154].

Changes in public attitudes and the lifting of COVID-19 restrictions due to the multifactorial desire to stimulate economic activities has required a shift of testing paradigms in 2022, despite warnings from public health officials against a hard exit from public health restrictions [155,156]. An important strategy for testing moving forward is to determine when someone becomes infectious or is no longer infectious following a positive test for COVID-19. Generally, patient specimens tend to not contain culturable virus past day 5 of symptom onset [157,158]. However, due to their sensitivity to post-infectious viral RNA in specimens, PCR-based methods may mislead individuals to believe that they are still infectious several days after symptom onset [131]. Furthermore, detection of viral RNA can occur days and weeks after an active infection due to the sensitivity of PCR-based methods [18,159,160].

In contrast, LFTs were thought to have poor sensitivity and their value for identifying infections and managing the pandemic was questioned [161,162]. However, LFTs do reliably detect SARS-CoV-2 proteins when there is a high viral load, which appears to correlate with a person's infectiousness [69,163]. Therefore, LFTs are an important diagnostic tool to determine infectiousness with fast turnaround times, ease of use, and accessibility by the general public [131,164]. One study has suggested that the test sensitivity of LFTs appears to be less important than accessibility to LFTs, frequent testing, and fast reporting times for reducing the impact of viral spread [165]. While PCR-based methods are important for COVID-19 surveillance, their use is labor intensive and time consuming, and laboratories are often slow to report results, rendering such methods limited in their surveillance capacity [131].

These limitations are demonstrated by the estimated 10-fold under-reporting of cases in the United States in 2020 due to shortages in testing and slow rollout of testing and slow reporting of results [166]. However, one strategy that may balance the strengths and weaknesses of both types of tests is to corroborate a positive LFT result using a PCR-based method. Indeed, in December 2021 sufficient surveillance and reduction of COVID-19 spread using this joint LFT-PCR strategy was demonstrated in Liverpool, U.K., where there was an estimated 21% reduction of cases [164,167].

# 8 What Lies Ahead

Diagnostic tools have played an important role during the COVID-19 pandemic. Different tests offer different advantages (Figure 1). Specifically, the results of SARS-CoV-2 diagnostic tests (typically qPCR or LFT-based tests) have been used to estimate the number of infections in the general population, thus informing public health strategies around the globe [26]. During the surges caused by the different SARS-CoV-2 variants between 2020 and 2021, government-sponsored efforts to conduct mass testing and to provide free diagnostic tests to the population were a common occurrence in many parts of the world [168,169,170]. However, recent reports indicate that such public health policies are starting to change during 2022. For example, it is known that the UK plans to dismantle its COVID-19 testing program and scale back its daily reporting requirements [171,172]. A similar approach can be seen in the US as well, where multiple state-run testing facilities are closing, despite some groups advocating to keep them open [173,174]. These ongoing changes in testing policy are likely to have a direct effect on how the pandemic is managed moving forward. SARS-CoV-2 diagnostic tests can be used effectively to slow the spread of the disease only when 1) they are used to share testing results in a timely manner so that they can reasonably be used to approximate the number of infections in the population and 2) those tests are easily accessible by the general public.

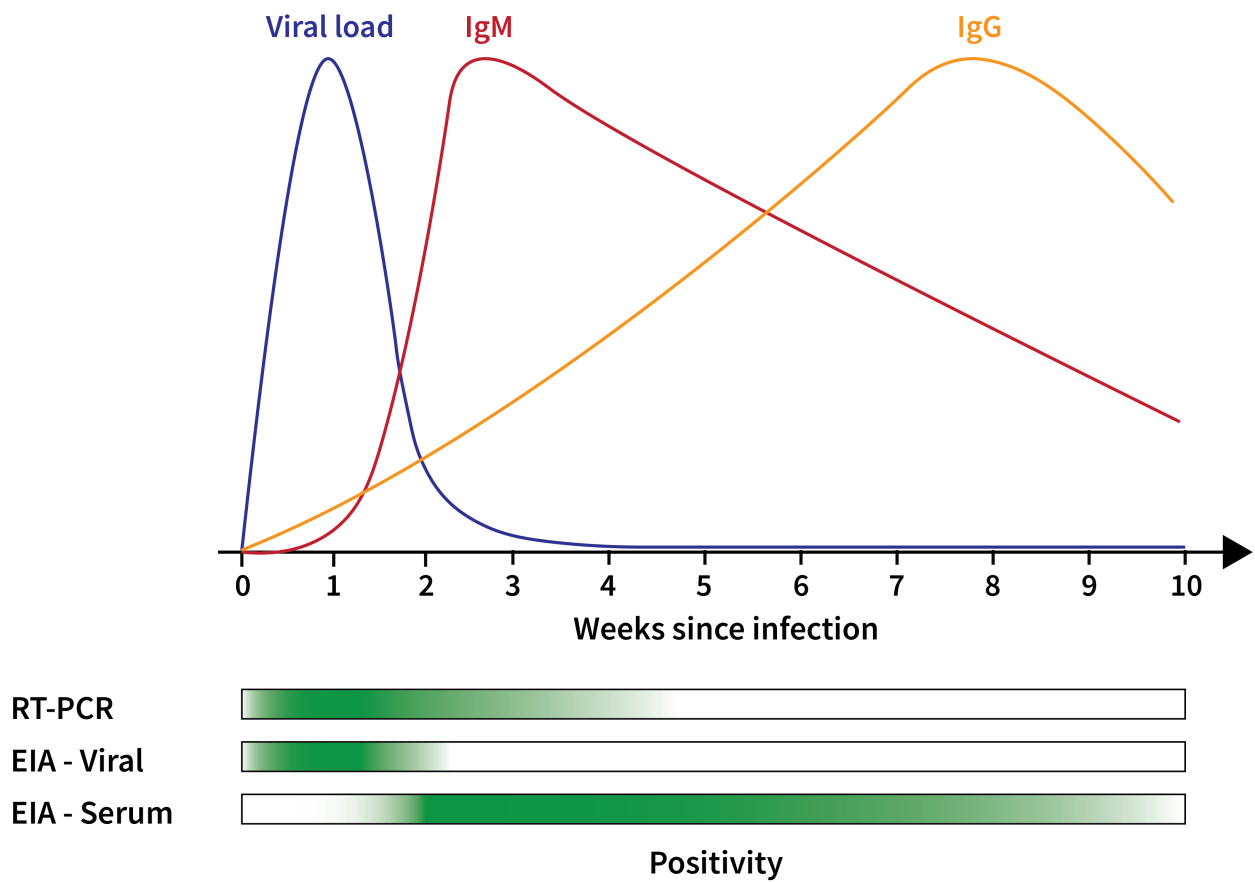

*Figure 1: **Summary of Diagnostic Technologies used in COVID-19 Testing.** The immune response to SARS-CoV-2 means that different diagnostic approaches offer different views of*

*COVID-19. Early in the infection course, viral load is high. This means that PCR-based testing and EIA testing for antigens are likely to return positives (as indicated by the green bars at the bottom). As viral load decreases, EIA antigen tests become negative, but PCR-based tests can still detect even very low viral loads. From a serological perspective, IgM peaks in the first few weeks following infection and then decreases, while IgG peaks much later in the infection course. Therefore, serological tests are likely to return positives in first few months following the acute infection course. Additional detail is available above and in several analyses and reviews [1,131,160,175].*

Children are one segment of the population where the importance of the two aforementioned conditions can be exemplified. This group is particularly vulnerable as there are ongoing challenges with testing in schools, increased COVID-19 mortality rates, and COVID-19-associated orphanhood. In this regard, although there is evidence of the efficacy of routine diagnostic testing to reduce the probability of having infectious students [176,177], as of March of 2022 there is an increasing number of schools that have stopped or plan to stop contact tracing efforts [178,179], in line with an announcement made by the CDC where it no longer recommended contact tracing as a strategy to contain the virus [180]. An estimated 197 children have died in the US from COVID-19 during the first three months of 2022 [181], compared to 735 deaths in the preceding 20 months of the pandemic [182], and millions of children have been orphaned as a consequence of parent or caregiver death due to COVID-19 [183]. It is likely that reducing or eliminating testing capacity in schools will directly exacerbate these negative outcomes for the remainder of 2022.

The SARS-CoV-2 diagnostic tools presented in this paper are far less useful if they are difficult to obtain, or if their limited use results in biased data that would lead to ill-informed public health strategies. Under conditions of limited supply, different strategies for testing are needed [184]. The pandemic is still an ongoing public health threat and it is worrying that active testing and tracing efforts are a low priority for public health authorities in many countries. If this trend continues, the lack of testing could result in increased morbidity and mortality and an overall failure to manage the pandemic.

# A  Additional Items

## A.1  Competing Interests

| Author | Competing Interests | Last Reviewed |
|---|---|---|
| Halie M. Rando | None | 2021-01-20 |
| Christian Brueffer | Employee and shareholder of SAGA Diagnostics AB. | 2020-11-11 |
| Ronan Lordan | None | 2020-11-03 |
| Anna Ada Dattoli | None | 2020-03-26 |
| David Manheim | None | 2022-03-15 |
| Jesse G. Meyer | None | 2022-01-06 |
| Ariel I. Mundo | None | 2021-12-19 |

| Author | Competing Interests | Last Reviewed |
|---|---|---|
| Dimitri Perrin | None | 2020-11-11 |
| David Mai | None | 2021-01-08 |
| Nils Wellhausen | None | 2020-11-03 |
| COVID-19 Review Consortium | None | 2021-01-16 |
| Anthony Gitter | Filed a patent application with the Wisconsin Alumni Research Foundation related to classifying activated T cells | 2020-11-10 |
| Casey S. Greene | None | 2021-01-20 |

## A.2 Author Contributions

| Author | Contributions |
|---|---|
| Halie M. Rando | Project Administration, Writing - Original Draft, Writing - Review & Editing, Visualization |
| Christian Brueffer | Project Administration, Writing - Original Draft, Writing - Review & Editing |
| Ronan Lordan | Project Administration, Writing - Original Draft, Writing - Review & Editing |
| Anna Ada Dattoli | Writing - Original Draft |
| David Manheim | Writing - Original Draft, Writing - Review & Editing |
| Jesse G. Meyer | Writing - Original Draft, Writing - Review & Editing |
| Ariel I. Mundo | Writing - Original Draft, Writing - Review & Editing |
| Dimitri Perrin | Writing - Original Draft, Writing - Review & Editing |
| David Mai | Writing - Original Draft, Writing - Review & Editing |
| Nils Wellhausen | Visualization, Writing - Review & Editing |
| COVID-19 Review Consortium | Project Administration |
| Anthony Gitter | Methodology, Project Administration, Software, Writing - Review & Editing |
| Casey S. Greene | Project Administration, Writing - Original Draft, Writing - Review & Editing, Conceptualization |

## A.3 Acknowledgements

We thank Yael Evelyn Marshall, who contributed writing (original draft) as well as reviewing and editing of pieces of the text but who did not formally approve the manuscript, as well as Ronnie Russell, who contributed text to and helped develop the structure of the manuscript early in the writing process. We also thank Matthias Fax, Joshua Blake, and consortium members Simina Boca, Lucy D'Agostino McGowan, and Greg Szeto, who helped with writing and editing text. We are also grateful to Nadia Danilova, James Eberwine and Ipsita Krishnan for feedback on early versions of the text.

# A.4 References

1. **Pathogenesis, Symptomatology, and Transmission of SARS-CoV-2 through Analysis of Viral Genomics and Structure** Halie M Rando, Adam L MacLean, Alexandra J Lee, Ronan Lordan, Sandipan Ray, Vikas Bansal, Ashwin N Skelly, Elizabeth Sell, John J Dziak, Lamonica Shinholster, … Casey S Greene *mSystems* (2021-10-26) https://pubmed.ncbi.nlm.nih.gov/34698547/ DOI: 10.1128/msystems.00095-21

2. **Aggressively find, test, trace and isolate to beat COVID-19** Larissa M Matukas, Irfan A Dhalla, Andreas Laupacis *Canadian Medical Association Journal* (2020-09-09) https://doi.org/gh2jvk DOI: 10.1503/cmaj.202120 · PMID: 32907821 · PMCID: PMC7546740

3. **COVID-19 in New Zealand and the impact of the national response: a descriptive epidemiological study** Sarah Jefferies, Nigel French, Charlotte Gilkison, Giles Graham, Virginia Hope, Jonathan Marshall, Caroline McElnay, Andrea McNeill, Petra Muellner, Shevaun Paine, … Patricia Priest *The Lancet Public Health* (2020-11) https://doi.org/ftzx DOI: 10.1016/s2468-2667(20)30225-5 · PMID: 33065023 · PMCID: PMC7553903

4. **Potential lessons from the Taiwan and New Zealand health responses to the COVID-19 pandemic** Jennifer Summers, Hao-Yuan Cheng, Hsien-Ho Lin, Lucy Telfar Barnard, Amanda Kvalsvig, Nick Wilson, Michael G Baker *The Lancet Regional Health - Western Pacific* (2020-11) https://doi.org/ghrzb4 DOI: 10.1016/j.lanwpc.2020.100044 · PMID: 34013216 · PMCID: PMC7577184

5. **Novel Coronavirus – China** https://www.who.int/emergencies/disease-outbreak-news/item/2020-DON233

6. **Detection of 2019 novel coronavirus (2019-nCoV) by real-time RT-PCR** Victor M Corman, Olfert Landt, Marco Kaiser, Richard Molenkamp, Adam Meijer, Daniel KW Chu, Tobias Bleicker, Sebastian Brünink, Julia Schneider, Marie Luisa Schmidt, … Christian Drosten *Eurosurveillance* (2020-01-23) https://doi.org/ggjs7g DOI: 10.2807/1560-7917.es.2020.25.3.2000045 · PMID: 31992387 · PMCID: PMC6988269

7. **Polymerase Chain Reaction** Lilit Garibyan, Nidhi Avashia *Journal of Investigative Dermatology* (2013-03) https://doi.org/ggtkkc DOI: 10.1038/jid.2013.1 · PMID: 23399825 · PMCID: PMC4102308

8. **Development and application of lateral flow test strip technology for detection of infectious agents and chemical contaminants: a review** Babacar Ngom, Yancheng Guo, Xiliang Wang, Dingren Bi *Analytical and Bioanalytical Chemistry* (2010-04-27) https://doi.org/cn8cn9 DOI: 10.1007/s00216-010-3661-4 · PMID: 20422164

9. **Enzyme Linked Immunosorbent Assay** Mandy Alhajj, Aisha Farhana *StatPearls* (2022-02-02) https://pubmed.ncbi.nlm.nih.gov/32310382 PMID: 32310382

10. **Evaluation and Comparison of Serological Methods for COVID-19 Diagnosis** Fanwu Gong, Hua-xing Wei, Qiangsheng Li, Liu Liu, Bofeng Li *Frontiers in Molecular*

*Biosciences* (2021-07-23) https://doi.org/gnxn82 DOI: 10.3389/fmolb.2021.682405 · PMID: 34368226 · PMCID: PMC8343015

11.     **Sensitivity in Detection of Antibodies to Nucleocapsid and Spike Proteins of Severe Acute Respiratory Syndrome Coronavirus 2 in Patients With Coronavirus Disease 2019** Peter D Burbelo, Francis X Riedo, Chihiro Morishima, Stephen Rawlings, Davey Smith, Sanchita Das, Jeffrey R Strich, Daniel S Chertow, Richard T Davey Jr, Jeffrey I Cohen *The Journal of Infectious Diseases* (2020-05-19) https://doi.org/ggxz2f DOI: 10.1093/infdis/jiaa273 · PMID: 32427334 · PMCID: PMC7313936

12.     **Temporal profiles of viral load in posterior oropharyngeal saliva samples and serum antibody responses during infection by SARS-CoV-2: an observational cohort study** Kelvin Kai-Wang To, Owen Tak-Yin Tsang, Wai-Shing Leung, Anthony Raymond Tam, Tak-Chiu Wu, David Christopher Lung, Cyril Chik-Yan Yip, Jian-Piao Cai, Jacky Man-Chun Chan, Thomas Shiu-Hong Chik, … Kwok-Yung Yuen *The Lancet Infectious Diseases* (2020-05) https://doi.org/ggp4qx DOI: 10.1016/s1473-3099(20)30196-1 · PMID: 32213337 · PMCID: PMC7158907

13.     **Fecal specimen diagnosis 2019 novel coronavirus–infected pneumonia** JingCheng Zhang, SaiBin Wang, YaDong Xue *Journal of Medical Virology* (2020-03-12) https://doi.org/ggpx6d DOI: 10.1002/jmv.25742 · PMID: 32124995

14.     **A beginner's guide to RT-PCR, qPCR and RT-qPCR** Grace Adams *The Biochemist* (2020-06-15) https://doi.org/gm8nfz DOI: 10.1042/bio20200034

15.     **The MIQE Guidelines: Minimum Information for Publication of Quantitative Real-Time PCR Experiments** Stephen A Bustin, Vladimir Benes, Jeremy A Garson, Jan Hellemans, Jim Huggett, Mikael Kubista, Reinhold Mueller, Tania Nolan, Michael W Pfaffl, Gregory L Shipley, … Carl T Wittwer *Clinical Chemistry* (2009-04-01) https://doi.org/fkkjq5 DOI: 10.1373/clinchem.2008.112797 · PMID: 19246619

16.     **Molecular Diagnosis of a Novel Coronavirus (2019-nCoV) Causing an Outbreak of Pneumonia** Daniel KW Chu, Yang Pan, Samuel MS Cheng, Kenrie PY Hui, Pavithra Krishnan, Yingzhi Liu, Daisy YM Ng, Carrie KC Wan, Peng Yang, Quanyi Wang, … Leo LM Poon *Clinical Chemistry* (2020-04) https://doi.org/ggnbpp DOI: 10.1093/clinchem/hvaa029 · PMID: 32031583 · PMCID: PMC7108203

17.     **Viral RNA load as determined by cell culture as a management tool for discharge of SARS-CoV-2 patients from infectious disease wards** Bernard La Scola, Marion Le Bideau, Julien Andreani, Van Thuan Hoang, Clio Grimaldier, Philippe Colson, Philippe Gautret, Didier Raoult *European Journal of Clinical Microbiology & Infectious Diseases* (2020-04-27) https://doi.org/ghf8cj DOI: 10.1007/s10096-020-03913-9 · PMID: 32342252 · PMCID: PMC7185831

18.     **Duration and key determinants of infectious virus shedding in hospitalized patients with coronavirus disease-2019 (COVID-19)** Jeroen JA van Kampen, David AMC van de Vijver, Pieter LA Fraaij, Bart L Haagmans, Mart M Lamers, Nisreen Okba, Johannes PC van den Akker, Henrik Endeman, Diederik AMPJ Gommers, Jan J Cornelissen, … Annemiek A van der Eijk *Nature Communications* (2021-01-11) https://doi.org/gjm8g2 DOI: 10.1038/s41467-020-20568-4 · PMID: 33431879 · PMCID: PMC7801729

19. **Duration of infectiousness and correlation with RT-PCR cycle threshold values in cases of COVID-19, England, January to May 2020** Anika Singanayagam, Monika Patel, Andre Charlett, Jamie Lopez Bernal, Vanessa Saliba, Joanna Ellis, Shamez Ladhani, Maria Zambon, Robin Gopal *Eurosurveillance* (2020-08-13) https://doi.org/gg9jt7 DOI: 10.2807/1560-7917.es.2020.25.32.2001483 · PMID: 32794447 · PMCID: PMC7427302

20. **Ct values from SARS-CoV-2 diagnostic PCR assays should not be used as direct estimates of viral load** Elias Dahdouh, Fernando Lázaro-Perona, María Pilar Romero-Gómez, Jesús Mingorance, Julio García-Rodriguez *Journal of Infection* (2021-03) https://doi.org/gjkr5s DOI: 10.1016/j.jinf.2020.10.017 · PMID: 33131699 · PMCID: PMC7585367

21. **Virological assessment of hospitalized patients with COVID-2019** Roman Wölfel, Victor M Corman, Wolfgang Guggemos, Michael Seilmaier, Sabine Zange, Marcel A Müller, Daniela Niemeyer, Terry C Jones, Patrick Vollmar, Camilla Rothe, … Clemens Wendtner *Nature* (2020-04-01) https://doi.org/ggqrv7 DOI: 10.1038/s41586-020-2196-x · PMID: 32235945

22. **Estimating the false-negative test probability of SARS-CoV-2 by RT-PCR** Paul S Wikramaratna, Robert S Paton, Mahan Ghafari, José Lourenço *Eurosurveillance* (2020-12-17) https://doi.org/gmb6n3 DOI: 10.2807/1560-7917.es.2020.25.50.2000568 · PMID: 33334398 · PMCID: PMC7812420

23. **Diagnosing SARS-CoV-2 infection: the danger of over-reliance on positive test results** Andrew N Cohen, Bruce Kessel, Michael G Milgroom *Cold Spring Harbor Laboratory* (2020-05-01) https://doi.org/gh3xk7 DOI: 10.1101/2020.04.26.20080911

24. **Diagnostic performance of different sampling approaches for SARS-CoV-2 RT-PCR testing: a systematic review and meta-analysis** Nicole Ngai Yung Tsang, Hau Chi So, Ka Yan Ng, Benjamin J Cowling, Gabriel M Leung, Dennis Kai Ming Ip *The Lancet Infectious Diseases* (2021-09) https://doi.org/gjrfns DOI: 10.1016/s1473-3099(21)00146-8 · PMID: 33857405 · PMCID: PMC8041361

25. **Improved Molecular Diagnosis of COVID-19 by the Novel, Highly Sensitive and Specific COVID-19-RdRp/Hel Real-Time Reverse Transcription-PCR Assay Validated In Vitro and with Clinical Specimens** Jasper Fuk-Woo Chan, Cyril Chik-Yan Yip, Kelvin Kai-Wang To, Tommy Hing-Cheung Tang, Sally Cheuk-Ying Wong, Kit-Hang Leung, Agnes Yim-Fong Fung, Anthony Chin-Ki Ng, Zijiao Zou, Hoi-Wah Tsoi, … Kwok-Yung Yuen *Journal of Clinical Microbiology* (2020-04-23) https://doi.org/ggpv64 DOI: 10.1128/jcm.00310-20 · PMID: 32132196 · PMCID: PMC7180250

26. **Testing at scale during the COVID-19 pandemic** Tim R Mercer, Marc Salit *Nature Reviews Genetics* (2021-05-04) https://doi.org/gjvw2n DOI: 10.1038/s41576-021-00360-w · PMID: 33948037 · PMCID: PMC8094986

27. **SARS-CoV-2 samples may escape detection because of a single point mutation in the N gene** Katharina Ziegler, Philipp Steininger, Renate Ziegler, Jörg Steinmann, Klaus Korn, Armin Ensser *Eurosurveillance* (2020-10-01) https://doi.org/ghnwss DOI: 10.2807/1560-7917.es.2020.25.39.2001650 · PMID: 33006300 · PMCID: PMC7531073

28. **Multiple assays in a real-time RT-PCR SARS-CoV-2 panel can mitigate the risk of loss of sensitivity by new genomic variants during the COVID-19 outbreak** Luis Peñarrubia, Maria Ruiz, Roberto Porco, Sonia N Rao, Martí Juanola-Falgarona, Davide Manissero, Marta López-Fontanals, Josep Pareja *International Journal of Infectious Diseases* (2020-08) https://doi.org/gmb6rv DOI: 10.1016/j.ijid.2020.06.027 · PMID: 32535302 · PMCID: PMC7289722

29. **Droplet Digital PCR versus qPCR for gene expression analysis with low abundant targets: from variable nonsense to publication quality data** Sean C Taylor, Genevieve Laperriere, Hugo Germain *Scientific Reports* (2017-05-25) https://doi.org/gbgsjw DOI: 10.1038/s41598-017-02217-x · PMID: 28546538 · PMCID: PMC5445070

30. **dPCR: A Technology Review** Phenix-Lan Quan, Martin Sauzade, Eric Brouzes *Sensors* (2018-04-20) https://doi.org/ggr39c DOI: 10.3390/s18041271 · PMID: 29677144 · PMCID: PMC5948698

31. **ddPCR: a more accurate tool for SARS-CoV-2 detection in low viral load specimens** Tao Suo, Xinjin Liu, Jiangpeng Feng, Ming Guo, Wenjia Hu, Dong Guo, Hafiz Ullah, Yang Yang, Qiuhan Zhang, Xin Wang, … Yu Chen *Emerging Microbes & Infections* (2020-06-07) https://doi.org/ggx2t2 DOI: 10.1080/22221751.2020.1772678 · PMID: 32438868 · PMCID: PMC7448897

32. **Highly accurate and sensitive diagnostic detection of SARS-CoV-2 by digital PCR** Lianhua Dong, Junbo Zhou, Chunyan Niu, Quanyi Wang, Yang Pan, Sitong Sheng, Xia Wang, Yongzhuo Zhang, Jiayi Yang, Manqing Liu, … Xiang Fang *Talanta* (2021-03) https://doi.org/gh2jvj DOI: 10.1016/j.talanta.2020.121726 · PMID: 33379001 · PMCID: PMC7588801

33. **Library preparation for next generation sequencing: A review of automation strategies** JF Hess, TA Kohl, M Kotrová, K Rönsch, T Paprotka, V Mohr, T Hutzenlaub, M Brüggemann, R Zengerle, S Niemann, N Paust *Biotechnology Advances* (2020-07) https://doi.org/ggth2v DOI: 10.1016/j.biotechadv.2020.107537 · PMID: 32199980

34. **Diagnosing COVID-19: The Disease and Tools for Detection** Buddhisha Udugama, Pranav Kadhiresan, Hannah N Kozlowski, Ayden Malekjahani, Matthew Osborne, Vanessa YC Li, Hongmin Chen, Samira Mubareka, Jonathan B Gubbay, Warren CW Chan *ACS Nano* (2020-03-30) https://doi.org/ggq8ds DOI: 10.1021/acsnano.0c02624 · PMID: 32223179 · PMCID: PMC7144809

35. **Emergence of a novel SARS-CoV-2 strain in Southern California, USA** Wenjuan Zhang, Brian D Davis, Stephanie S Chen, Jorge MSincuir Martinez, Jasmine T Plummer, Eric

Vail *Cold Spring Harbor Laboratory* (2021-01-20) https://doi.org/ghvq48 DOI: 10.1101/2021.01.18.21249786

36. **Genomic surveillance to combat COVID-19: challenges and opportunities** Janet D Robishaw, Scott M Alter, Joshua J Solano, Richard D Shih, David L DeMets, Dennis G Maki, Charles H Hennekens *The Lancet Microbe* (2021-09) https://doi.org/hkwr DOI: 10.1016/s2666-5247(21)00121-x · PMID: 34337584 · PMCID: PMC8315763

37. **Cov-Lineages** https://cov-lineages.org/

38. **Global disparities in SARS-CoV-2 genomic surveillance** Anderson F Brito, Elizaveta Semenova, Gytis Dudas, Gabriel W Hassler, Chaney C Kalinich, Moritz UG Kraemer, Joses Ho, Houriiyah Tegally, George Githinji, Charles N Agoti, … *Cold Spring Harbor Laboratory* (2021-08-26) https://doi.org/gn2tht DOI: 10.1101/2021.08.21.21262393 · PMID: 34462754 · PMCID: PMC8404891

39. **Evaluation of COVID-19 RT-qPCR Test in Multi sample Pools** Idan Yelin, Noga Aharony, Einat Shaer Tamar, Amir Argoetti, Esther Messer, Dina Berenbaum, Einat Shafran, Areen Kuzli, Nagham Gandali, Omer Shkedi, … Roy Kishony *Clinical Infectious Diseases* (2020-05-02) https://doi.org/ggtx9r DOI: 10.1093/cid/ciaa531 · PMID: 32358960 · PMCID: PMC7197588

40. **Analytical Validation of a COVID-19 qRT-PCR Detection Assay Using a 384-well Format and Three Extraction Methods** Andrew C Nelson, Benjamin Auch, Matthew Schomaker, Daryl M Gohl, Patrick Grady, Darrell Johnson, Robyn Kincaid, Kylene E Karnuth, Jerry Daniel, Jessica K Fiege, … Sophia Yohe *Cold Spring Harbor Laboratory* (2020-04-05) https://doi.org/ggs45d DOI: 10.1101/2020.04.02.022186

41. **Two-Stage Adaptive Pooling with RT-QPCR for Covid-19 Screening** Anoosheh Heidarzadeh, Krishna Narayanan *ICASSP 2021 - 2021 IEEE International Conference on Acoustics, Speech and Signal Processing (ICASSP)* (2021-06-06) https://doi.org/gppbs9 DOI: 10.1109/icassp39728.2021.9413685

42. **Pooled RT-qPCR testing for SARS-CoV-2 surveillance in schools - a cluster randomised trial** Alexander Joachim, Felix Dewald, Isabelle Suárez, Michael Zemlin, Isabelle Lang, Regine Stutz, Anna Marthaler, Hans Martin Bosse, Nadine Lübke, Juliane Münch, … Anna Kern *EClinicalMedicine* (2021-09) https://doi.org/gpzfjz DOI: 10.1016/j.eclinm.2021.101082 · PMID: 34458708 · PMCID: PMC8384501

43. **Pooling RT-qPCR testing for SARS-CoV-2 in 1000 individuals of healthy and infection-suspected patients** Yosuke Hirotsu, Makoto Maejima, Masahiro Shibusawa, Yuki Nagakubo, Kazuhiro Hosaka, Kenji Amemiya, Hitomi Sueki, Miyoko Hayakawa, Hitoshi Mochizuki, Toshiharu Tsutsui, … Masao Omata *Scientific Reports* (2020-11-03) https://doi.org/gpzfj4 DOI: 10.1038/s41598-020-76043-z · PMID: 33144632 · PMCID: PMC7641135

44.     **Loop-mediated isothermal amplification of DNA** T Notomi *Nucleic Acids Research* (2000-06-15) https://doi.org/bx567n DOI: 10.1093/nar/28.12.e63 · PMID: 10871386 · PMCID: PMC102748

45.     **A molecular test based on RT-LAMP for rapid, sensitive and inexpensive colorimetric detection of SARS-CoV-2 in clinical samples** Catarina Amaral, Wilson Antunes, Elin Moe, Américo G Duarte, Luís MP Lima, Cristiana Santos, Inês L Gomes, Gonçalo S Afonso, Ricardo Vieira, Helena Sofia S Teles, … Catarina Pimentel *Scientific Reports* (2021-08-12) https://doi.org/gnx4h3 DOI: 10.1038/s41598-021-95799-6 · PMID: 34385527 · PMCID: PMC8361189

46.     **Can the cycle threshold (Ct) value of RT-PCR test for SARS CoV2 predict infectivity among close contacts?** Soha Al Bayat, Jesha Mundodan, Samina Hasnain, Mohamed Sallam, Hayat Khogali, Dina Ali, Saif Alateeg, Mohamed Osama, Aiman Elberdiny, Hamad Al-Romaihi, Mohammed Hamad J Al-Thani *Journal of Infection and Public Health* (2021-09) https://doi.org/gnx4jw DOI: 10.1016/j.jiph.2021.08.013 · PMID: 34416598 · PMCID: PMC8362640

47.     **Evaluation of cycle threshold values at deisolation** Clayton T Mowrer, Hannah Creager, Kelly Cawcutt, Justin Birge, Elizabeth Lyden, Trevor C Van Schooneveld, Mark E Rupp, Angela Hewlett *Infection Control & Hospital Epidemiology* (2021-04-06) https://doi.org/gnx4jx DOI: 10.1017/ice.2021.132 · PMID: 33820588 · PMCID: PMC8060537

48.     **To Interpret the SARS-CoV-2 Test, Consider the Cycle Threshold Value** Michael R Tom, Michael J Mina *Clinical Infectious Diseases* (2020-05-21) https://doi.org/ggxf2s DOI: 10.1093/cid/ciaa619 · PMID: 32435816 · PMCID: PMC7314112

49.     **Rapid Molecular Detection of SARS-CoV-2 (COVID-19) Virus RNA Using Colorimetric LAMP** Yinhua Zhang, Nelson Odiwuor, Jin Xiong, Luo Sun, Raphael Ohuru Nyaruaba, Hongping Wei, Nathan A Tanner *Cold Spring Harbor Laboratory* (2020-02-29) https://doi.org/ggx3wn DOI: 10.1101/2020.02.26.20028373

50.     **Reverse Transcriptase Loop Mediated Isothermal Amplification (RT-LAMP) for COVID-19 diagnosis: a systematic review and meta-analysis.** Anita Dominique Subali, Lowilius Wiyono *Pathogens and global health* (2021-06-04) https://www.ncbi.nlm.nih.gov/pubmed/34086539 DOI: 10.1080/20477724.2021.1933335 · PMID: 34086539 · PMCID: PMC8182821

51.     **The CRISPR tool kit for genome editing and beyond** Mazhar Adli *Nature Communications* (2018-05-15) https://doi.org/gdj266 DOI: 10.1038/s41467-018-04252-2 · PMID: 29765029 · PMCID: PMC5953931

52.     **Nucleic acid detection with CRISPR-Cas13a/C2c2** Jonathan S Gootenberg, Omar O Abudayyeh, Jeong Wook Lee, Patrick Essletzbichler, Aaron J Dy, Julia Joung, Vanessa Verdine, Nina Donghia, Nichole M Daringer, Catherine A Freije, … Feng Zhang *Science* (2017-04-28) https://doi.org/f93x8p DOI: 10.1126/science.aam9321 · PMID: 28408723 · PMCID: PMC5526198

53.	**Development and evaluation of a rapid CRISPR-based diagnostic for COVID-19** Tieying Hou, Weiqi Zeng, Minling Yang, Wenjing Chen, Lili Ren, Jingwen Ai, Ji Wu, Yalong Liao, Xuejing Gou, Yongjun Li, … Teng Xu *PLOS Pathogens* (2020-08-27) https://doi.org/ghn7rp DOI: 10.1371/journal.ppat.1008705 · PMID: 32853291 · PMCID: PMC7451577

54.	**CRISPR-based surveillance for COVID-19 using genomically-comprehensive machine learning design** Hayden C Metsky, Catherine A Freije, Tinna-Solveig F Kosoko-Thoroddsen, Pardis C Sabeti, Cameron Myhrvold *Cold Spring Harbor Laboratory* (2020-03-02) https://doi.org/ggr3zf DOI: 10.1101/2020.02.26.967026

55.	**A Scalable, Easy-to-Deploy Protocol for Cas13-Based Detection of SARS-CoV-2 Genetic Material** Jennifer N Rauch, Eric Valois, Sabrina C Solley, Friederike Braig, Ryan S Lach, Morgane Audouard, Jose Carlos Ponce-Rojas, Michael S Costello, Naomi J Baxter, Kenneth S Kosik, … Maxwell Z Wilson *Journal of Clinical Microbiology* (2021-03-19) https://doi.org/gh3nm3 DOI: 10.1128/jcm.02402-20 · PMID: 33478979 · PMCID: PMC8092748

56.	**CRISPR–Cas12-based detection of SARS-CoV-2** James P Broughton, Xianding Deng, Guixia Yu, Clare L Fasching, Venice Servellita, Jasmeet Singh, Xin Miao, Jessica A Streithorst, Andrea Granados, Alicia Sotomayor-Gonzalez, … Charles Y Chiu *Nature Biotechnology* (2020-04-16) https://doi.org/ggv47f DOI: 10.1038/s41587-020-0513-4 · PMID: 32300245

57.	**An ultrasensitive, rapid, and portable coronavirus SARS-CoV-2 sequence detection method based on CRISPR-Cas12** Curti Lucia, Pereyra-Bonnet Federico, Gimenez Carla Alejandra *Cold Spring Harbor Laboratory* (2020-03-02) https://doi.org/gg7km6 DOI: 10.1101/2020.02.29.971127

58.	**Rapid detection of SARS-CoV-2 with CRISPR-Cas12a** Dan Xiong, Wenjun Dai, Jiaojiao Gong, Guande Li, Nansong Liu, Wei Wu, Jiaqiang Pan, Chen Chen, Yingzhen Jiao, Huina Deng, … Guanghui Tang *PLOS Biology* (2020-12-15) https://doi.org/gh3nm6 DOI: 10.1371/journal.pbio.3000978 · PMID: 33320883 · PMCID: PMC7737895

59.	**Ultrasensitive and visual detection of SARS-CoV-2 using all-in-one dual CRISPR-Cas12a assay** Xiong Ding, Kun Yin, Ziyue Li, Rajesh V Lalla, Enrique Ballesteros, Maroun M Sfeir, Changchun Liu *Nature Communications* (2020-09-18) https://doi.org/ghwjb2 DOI: 10.1038/s41467-020-18575-6 · PMID: 32948757 · PMCID: PMC7501862

60.	**SARS-CoV-2 detection with CRISPR diagnostics** Lu Guo, Xuehan Sun, Xinge Wang, Chen Liang, Haiping Jiang, Qingqin Gao, Moyu Dai, Bin Qu, Sen Fang, Yihuan Mao, … Wei Li *Cell Discovery* (2020-05-19) https://doi.org/ggx2wt DOI: 10.1038/s41421-020-0174-y · PMID: 32435508 · PMCID: PMC7235268

61.	**Electric field-driven microfluidics for rapid CRISPR-based diagnostics and its application to detection of SARS-CoV-2** Ashwin Ramachandran, Diego A Huyke, Eesha Sharma, Malaya K Sahoo, ChunHong Huang, Niaz Banaei, Benjamin A Pinsky, Juan G Santiago *Proceedings of the National Academy of Sciences* (2020-11-04)

https://doi.org/gh3nms DOI: 10.1073/pnas.2010254117 · PMID: 33148808 · PMCID: PMC7703567

62. **Annotation and Classification of CRISPR-Cas Systems** Kira S Makarova, Eugene V Koonin *Methods in Molecular Biology* (2015) https://doi.org/gppg52 DOI: 10.1007/978-1-4939-2687-9_4 · PMID: 25981466 · PMCID: PMC5901762

63. **Type III CRISPR-Cas systems: when DNA cleavage just isn't enough** Nora C Pyenson, Luciano A Marraffini *Current Opinion in Microbiology* (2017-06) https://doi.org/gcx7r3 DOI: 10.1016/j.mib.2017.08.003 · PMID: 28865392

64. **Type III CRISPR-Cas Immunity: Major Differences Brushed Aside** Gintautas Tamulaitis, Česlovas Venclovas, Virginijus Siksnys *Trends in Microbiology* (2017-01) https://doi.org/f9nj96 DOI: 10.1016/j.tim.2016.09.012 · PMID: 27773522

65. **Intrinsic signal amplification by type III CRISPR-Cas systems provides a sequence-specific SARS-CoV-2 diagnostic** Andrew Santiago-Frangos, Laina N Hall, Anna Nemudraia, Artem Nemudryi, Pushya Krishna, Tanner Wiegand, Royce A Wilkinson, Deann T Snyder, Jodi F Hedges, Calvin Cicha, … Blake Wiedenheft *Cell Reports Medicine* (2021-06) https://doi.org/gpmv6n DOI: 10.1016/j.xcrm.2021.100319 · PMID: 34075364 · PMCID: PMC8157118

66. **Amplification-free detection of SARS-CoV-2 with CRISPR-Cas13a and mobile phone microscopy** Parinaz Fozouni, Sungmin Son, María Díaz de León Derby, Gavin J Knott, Carley N Gray, Michael V D'Ambrosio, Chunyu Zhao, Neil A Switz, GRenuka Kumar, Stephanie I Stephens, … Melanie Ott *Cell* (2021-01) https://doi.org/ghnszx DOI: 10.1016/j.cell.2020.12.001 · PMID: 33306959 · PMCID: PMC7834310

67. **Rapid, Sensitive, and Specific Severe Acute Respiratory Syndrome Coronavirus 2 Detection: A Multicenter Comparison Between Standard Quantitative Reverse-Transcriptase Polymerase Chain Reaction and CRISPR-Based DETECTR** Eelke Brandsma, Han JMP Verhagen, Thijs JW van de Laar, Eric CJ Claas, Marion Cornelissen, Emile van den Akker *The Journal of Infectious Diseases* (2020-10-10) https://doi.org/gh3nmt DOI: 10.1093/infdis/jiaa641 · PMID: 33535237 · PMCID: PMC7665660

68. **Rapid, accurate, nucleobase detection using FnCas9** Mohd Azhar, Rhythm Phutela, Manoj Kumar, Asgar Hussain Ansari, Riya Rauthan, Sneha Gulati, Namrata Sharma, Dipanjali Sinha, Saumya Sharma, Sunaina Singh, … *Cold Spring Harbor Laboratory* (2020-09-14) https://doi.org/gh3nmv DOI: 10.1101/2020.09.13.20193581

69. **Rapid Diagnostic Testing for SARS-CoV-2** Paul K Drain *New England Journal of Medicine* (2022-01-20) https://doi.org/gn2sfk DOI: 10.1056/nejmcp2117115 · PMID: 34995029 · PMCID: PMC8820190

70. **The Performance of Two Rapid Antigen Tests During Population-Level Screening for SARS-CoV-2 Infection** Mohammad Alghounaim, Hamad Bastaki, Farah Bin Essa, Hoda

Motlagh, Salman Al-Sabah *Frontiers in Medicine* (2021-12-23) https://doi.org/gpp4xw DOI: 10.3389/fmed.2021.797109 · PMID: 35004772 · PMCID: PMC8733308

71. **Lateral flow (immuno)assay: its strengths, weaknesses, opportunities and threats. A literature survey** Geertruida A Posthuma-Trumpie, Jakob Korf, Aart van Amerongen *Analytical and Bioanalytical Chemistry* (2008-08-13) https://doi.org/bcsjdw DOI: 10.1007/s00216-008-2287-2 · PMID: 18696055

72. **A systematic review of the sensitivity and specificity of lateral flow devices in the detection of SARS-CoV-2.** Dylan A Mistry, Jenny Y Wang, Mika-Erik Moeser, Thomas Starkey, Lennard YW Lee *BMC infectious diseases* (2021-08-18) https://www.ncbi.nlm.nih.gov/pubmed/34407759 DOI: 10.1186/s12879-021-06528-3 · PMID: 34407759 · PMCID: PMC8371300

73. **Government sets out next steps for living with COVID** GOV.UK https://www.gov.uk/government/news/government-sets-out-next-steps-for-living-with-covid

74. **Fact Sheet: The Biden Administration to Begin Distributing At-Home, Rapid COVID-19 Tests to Americans for Free** The White House (2022-01-14) https://www.whitehouse.gov/briefing-room/statements-releases/2022/01/14/fact-sheet-the-biden-administration-to-begin-distributing-at-home-rapid-covid-19-tests-to-americans-for-free/

75. **Scaling up COVID-19 rapid antigen tests: promises and challenges** Rosanna W Peeling, Piero L Olliaro, Debrah I Boeras, Noah Fongwen *The Lancet Infectious Diseases* (2021-09) https://doi.org/hk34 DOI: 10.1016/s1473-3099(21)00048-7 · PMID: 33636148 · PMCID: PMC7906660

76. **Detection technologies and recent developments in the diagnosis of COVID-19 infection** Praveen Rai, Ballamoole Krishna Kumar, Vijaya Kumar Deekshit, Indrani Karunasagar, Iddya Karunasagar *Applied Microbiology and Biotechnology* (2021-01) https://doi.org/gnvtp9 DOI: 10.1007/s00253-020-11061-5 · PMID: 33394144 · PMCID: PMC7780074

77. **Enzyme Immunoassay (EIA)/Enzyme-Linked Immunosorbent Assay (ELISA)** Rudolf M Lequin *Clinical Chemistry* (2005-12-01) https://doi.org/dts5xp DOI: 10.1373/clinchem.2005.051532 · PMID: 16179424

78. **Enzyme-Immunoassay: A Powerful Analytical Tool** AHWM Schuurs, BK Van Weemen *Journal of Immunoassay* (1980-01) https://doi.org/btqbvh DOI: 10.1080/01971528008055786 · PMID: 6785317

79. **Antibody testing for COVID-19: A report from the National COVID Scientific Advisory Panel** Emily R Adams, Mark Ainsworth, Rekha Anand, Monique I Andersson, Kathryn Auckland, JKenneth Baillie, Eleanor Barnes, Sally Beer, John I Bell, Tamsin Berry, … *Wellcome Open Research* (2020-06-11) https://doi.org/gpp4wq DOI: 10.12688/wellcomeopenres.15927.1 · PMID: 33748431 · PMCID: PMC7941096

80. **Detection of SARS-CoV-2 by antigen ELISA test is highly swayed by viral load and sample storage condition** Nihad Adnan, Shahad Saif Khandker, Ahsanul Haq, Mousumi Akter Chaity, Abdul Khalek, Anawarul Quader Nazim, Taku Kaitsuka, Kazuhito Tomizawa, Masayasu Mie, Eiry Kobatake, … MohdRaeed Jamiruddin *Expert Review of Anti-infective Therapy* (2021-09-11) https://doi.org/gprjx9 DOI: 10.1080/14787210.2021.1976144 · PMID: 34477019 · PMCID: PMC8442762

81. **Negative Nasopharyngeal and Oropharyngeal Swabs Do Not Rule Out COVID-19** Poramed Winichakoon, Romanee Chaiwarith, Chalerm Liwsrisakun, Parichat Salee, Aree Goonna, Atikun Limsukon, Quanhathai Kaewpoowat *Journal of Clinical Microbiology* (2020-04-23) https://doi.org/ggpw9m DOI: 10.1128/jcm.00297-20 · PMID: 32102856 · PMCID: PMC7180262

82. **Coronavirus and the race to distribute reliable diagnostics** Cormac Sheridan *Nature Biotechnology* (2020-02-19) https://doi.org/ggm4nt DOI: 10.1038/d41587-020-00002-2 · PMID: 32265548

83. **Quarantine & Isolation** CDC *Centers for Disease Control and Prevention* (2022-03-30) https://www.cdc.gov/coronavirus/2019-ncov/your-health/quarantine-isolation.html

84. **CRISPR-Cas System: An Approach With Potentials for COVID-19 Diagnosis and Therapeutics** Prashant Kumar, Yashpal Singh Malik, Balasubramanian Ganesh, Somnath Rahangdale, Sharad Saurabh, Senthilkumar Natesan, Ashish Srivastava, Khan Sharun, MohdIqbal Yatoo, Ruchi Tiwari, … Kuldeep Dhama *Frontiers in Cellular and Infection Microbiology* (2020-11-02) https://doi.org/ghz57p DOI: 10.3389/fcimb.2020.576875 · PMID: 33251158 · PMCID: PMC7673385

85. **The standard coronavirus test, if available, works well—but can new diagnostics help in this pandemic?** Robert Service *Science* (2020-03-22) https://doi.org/ggq9wm DOI: 10.1126/science.abb8400

86. **Laboratory Diagnosis of COVID-19: Current Issues and Challenges** Yi-Wei Tang, Jonathan E Schmitz, David H Persing, Charles W Stratton *Journal of Clinical Microbiology* (2020-05-26) https://doi.org/ggq7h8 DOI: 10.1128/jcm.00512-20 · PMID: 32245835

87. **The Plane Is Boarding, Where Are Your Test Results?** Lauren Sloss *The New York Times* (2021-12-31) https://www.nytimes.com/2021/12/31/travel/covid-test-chaos.html

88. **A Stark Contrast Between the U.S. and Europe on Tests** Naomi Kresge *Bloomberg* (2022-01-07) https://www.bloomberg.com/news/newsletters/2022-01-07/a-stark-contrast-between-the-u-s-and-europe-on-tests

89. **IgG Subclasses and Allotypes: From Structure to Effector Functions** Gestur Vidarsson, Gillian Dekkers, Theo Rispens *Frontiers in Immunology* (2014-10-20) https://doi.org/gc6vx6 DOI: 10.3389/fimmu.2014.00520 · PMID: 25368619 · PMCID: PMC4202688

90. **The distribution and functions of immunoglobulin isotypes** Charles A Janeway Jr, Paul Travers, Mark Walport, Mark J Shlomchik (editors) *Immunobiology: The Immune System in Health and Disease* (2001) https://www.ncbi.nlm.nih.gov/books/NBK27162 ISBN: 978-0815336426

91. **Longitudinal profile of antibodies against SARS-coronavirus in SARS patients and their clinical significance** Hongying MO, Guangqiao ZENG, Xiaolan REN, Hui LI, Changwen KE, Yaxia TAN, Chaoda CAI, Kefang LAI, Rongchang CHEN, Moira CHAN-YEUNG, Nanshan ZHONG *Respirology* (2006-01) https://doi.org/dn23vj DOI: 10.1111/j.1440-1843.2006.00783.x · PMID: 16423201 · PMCID: PMC7192223

92. **Detection of antibodies against SARS-CoV-2 in patients with COVID-19** Zhe Du, Fengxue Zhu, Fuzheng Guo, Bo Yang, Tianbing Wang *Journal of Medical Virology* (2020-04-10) https://doi.org/ggq7m2 DOI: 10.1002/jmv.25820 · PMID: 32243608

93. **Review of Current Advances in Serologic Testing for COVID-19** Andrea P Espejo, Yamac Akgun, Abdulaziz F Al Mana, Youley Tjendra, Nicolas C Millan, Carmen Gomez-Fernandez, Carolyn Cray *American Journal of Clinical Pathology* (2020-06-25) https://doi.org/gkzfrn DOI: 10.1093/ajcp/aqaa112 · PMID: 32583852 · PMCID: PMC7337672

94. **A serological assay to detect SARS-CoV-2 seroconversion in humans** Fatima Amanat, Daniel Stadlbauer, Shirin Strohmeier, Thi HO Nguyen, Veronika Chromikova, Meagan McMahon, Kaijun Jiang, Guha Asthagiri Arunkumar, Denise Jurczyszak, Jose Polanco, … Florian Krammer *Nature Medicine* (2020-05-12) https://doi.org/ggx28b DOI: 10.1038/s41591-020-0913-5 · PMID: 32398876 · PMCID: PMC8183627

95. **Development of a Fast SARS-CoV-2 IgG ELISA, Based on Receptor-Binding Domain, and Its Comparative Evaluation Using Temporally Segregated Samples From RT-PCR Positive Individuals** Farha Mehdi, Souvick Chattopadhyay, Ramachandran Thiruvengadam, Sarla Yadav, Manjit Kumar, Sangita Kumari Sinha, Sandeep Goswami, Pallavi Kshetrapal, Nitya Wadhwa, Uma Chandramouli Natchu, … Gaurav Batra *Frontiers in Microbiology* (2021-01-20) https://doi.org/gpp5md DOI: 10.3389/fmicb.2020.618097 · PMID: 33552028 · PMCID: PMC7854536

96. **Development and evaluation of a low cost IgG ELISA test based in RBD protein for COVID-19** Luciana Villafañe, Lucía Gallo Vaulet, Florencia M Viere, Laura I Klepp, Marina A Forrellad, María M Bigi, María I Romano, Giovanni Magistrelli, Marcelo Rodríguez Fermepin, Fabiana Bigi *Journal of Immunological Methods* (2022-01) https://doi.org/gpp4wh DOI: 10.1016/j.jim.2021.113182 · PMID: 34762914 · PMCID: PMC8574101

97. **A sensitive and rapid chemiluminescence immunoassay for point-of-care testing (POCT) of copeptin in serum based on high-affinity monoclonal antibodies via cytokine-assisted immunization** Yu Wang, Emmanuel Enoch Dzakah, Ye Kang, Yanxue Cai, Peidian Wu, Bo Tang, Run Li, Xiaowei He *International Journal of Nanomedicine* (2019-06) https://doi.org/gpp5mc DOI: 10.2147/ijn.s200556 · PMID: 31354261 · PMCID: PMC6580123

98. **Chemiluminescent immunoassay technology: what does it change in autoantibody detection?** Luigi Cinquanta, Desré Ethel Fontana, Nicola Bizzaro *Autoimmunity Highlights* (2017-06-24) https://doi.org/gh6hcm DOI: 10.1007/s13317-017-0097-2 · PMID: 28647912 · PMCID: PMC5483212

99. **A Peptide-Based Magnetic Chemiluminescence Enzyme Immunoassay for Serological Diagnosis of Coronavirus Disease 2019** Xue-fei Cai, Juan Chen, Jie-li Hu, Quan-xin Long, Hai-jun Deng, Ping Liu, Kai Fan, Pu Liao, Bei-zhong Liu, Gui-cheng Wu, … De-qiang Wang *The Journal of Infectious Diseases* (2020-07-15) https://doi.org/ggv2fx DOI: 10.1093/infdis/jiaa243 · PMID: 32382737 · PMCID: PMC7239108

100. **Comparison of SARS-CoV-2 serological tests with different antigen targets** Alix T Coste, Katia Jaton, Matthaios Papadimitriou-Olivgeris, Gilbert Greub, Antony Croxatto *Journal of Clinical Virology* (2021-01) https://doi.org/gk8s5q DOI: 10.1016/j.jcv.2020.104690 · PMID: 33253926 · PMCID: PMC7670982

101. **Assessment of SARS-CoV-2 serological tests for the diagnosis of COVID-19 through the evaluation of three immunoassays: Two automated immunoassays (Euroimmun and Abbott) and one rapid lateral flow immunoassay (NG Biotech)** Thomas Nicol, Caroline Lefeuvre, Orianne Serri, Adeline Pivert, Françoise Joubaud, Vincent Dubée, Achille Kouatchet, Alexandra Ducancelle, Françoise Lunel-Fabiani, Hélène Le Guillou-Guillemette *Journal of Clinical Virology* (2020-08) https://doi.org/gg2ks6 DOI: 10.1016/j.jcv.2020.104511 · PMID: 32593133 · PMCID: PMC7295485

102. **Evaluation of Six Commercial Mid- to High-Volume Antibody and Six Point-of-Care Lateral Flow Assays for Detection of SARS-CoV-2 Antibodies** Carmen L Charlton, Jamil N Kanji, Kam Johal, Ashley Bailey, Sabrina S Plitt, Clayton MacDonald, Andrea Kunst, Emily Buss, Laura E Burnes, Kevin Fonseca, … Graham Tipples *Journal of Clinical Microbiology* (2020-09-22) https://doi.org/gpp5mb DOI: 10.1128/jcm.01361-20 · PMID: 32665420 · PMCID: PMC7512179

103. **Diagnostic accuracy of an automated chemiluminescent immunoassay for anti-SARS-CoV-2 IgM and IgG antibodies: an Italian experience** Maria Infantino, Valentina Grossi, Barbara Lari, Riccardo Bambi, Alessandro Perri, Matteo Manneschi, Giovanni Terenzi, Irene Liotti, Giovanni Ciotta, Cristina Taddei, … Mariangela Manfredi *Journal of Medical Virology* (2020-05-10) https://doi.org/ggv4c6 DOI: 10.1002/jmv.25932 · PMID: 32330291 · PMCID: PMC7264663

104. **SARS-CoV-2 serology: Validation of high-throughput chemiluminescent immunoassay (CLIA) platforms and a field study in British Columbia** Inna Sekirov, Vilte E Barakauskas, Janet Simons, Darrel Cook, Brandon Bates, Laura Burns, Shazia Masud, Marthe Charles, Meghan McLennan, Annie Mak, … Muhammad Morshed *Journal of Clinical Virology* (2021-09) https://doi.org/gpp5kb DOI: 10.1016/j.jcv.2021.104914 · PMID: 34304088 · PMCID: PMC8282439

105. **Cellex qSARS-CoV-2 IgG/IgM Rapid Test** Cellex (2020-04-07) https://www.fda.gov/media/136625/download

106. **Evaluation of Humoral Immune Response after SARS-CoV-2 Vaccination Using Two Binding Antibody Assays and a Neutralizing Antibody Assay** Minjeong Nam, Jong Do Seo, Hee-Won Moon, Hanah Kim, Mina Hur, Yeo-Min Yun *Microbiology Spectrum* (2021-12-22) https://doi.org/gnkz5s DOI: 10.1128/spectrum.01202-21 · PMID: 34817223 · PMCID: PMC8612149

107. **A high-throughput neutralizing antibody assay for COVID-19 diagnosis and vaccine evaluation** Antonio E Muruato, Camila R Fontes-Garfias, Ping Ren, Mariano A Garcia-Blanco, Vineet D Menachery, Xuping Xie, Pei-Yong Shi *Nature Communications* (2020-08-13) https://doi.org/gjkvr9 DOI: 10.1038/s41467-020-17892-0 · PMID: 32792628 · PMCID: PMC7426916

108. **Neutralizing antibody levels are highly predictive of immune protection from symptomatic SARS-CoV-2 infection** David S Khoury, Deborah Cromer, Arnold Reynaldi, Timothy E Schlub, Adam K Wheatley, Jennifer A Juno, Kanta Subbarao, Stephen J Kent, James A Triccas, Miles P Davenport *Nature Medicine* (2021-05-17) https://doi.org/gj3h47 DOI: 10.1038/s41591-021-01377-8 · PMID: 34002089

109. **Neutralizing Antibodies Correlate with Protection from SARS-CoV-2 in Humans during a Fishery Vessel Outbreak with a High Attack Rate** Amin Addetia, Katharine HD Crawford, Adam Dingens, Haiying Zhu, Pavitra Roychoudhury, Meei-Li Huang, Keith R Jerome, Jesse D Bloom, Alexander L Greninger *Journal of Clinical Microbiology* (2020-10-21) https://doi.org/gk7n4d DOI: 10.1128/jcm.02107-20 · PMID: 32826322 · PMCID: PMC7587101

110. **Neutralising antibody titres as predictors of protection against SARS-CoV-2 variants and the impact of boosting: a meta-analysis** Deborah Cromer, Megan Steain, Arnold Reynaldi, Timothy E Schlub, Adam K Wheatley, Jennifer A Juno, Stephen J Kent, James A Triccas, David S Khoury, Miles P Davenport *The Lancet Microbe* (2022-01) https://doi.org/gnx76z DOI: 10.1016/s2666-5247(21)00267-6 · PMID: 34806056 · PMCID: PMC8592563

111. **A cell-free high throughput assay for assessment of SARS-CoV-2 neutralizing antibodies** Sara Mravinacova, Malin Jönsson, Wanda Christ, Jonas Klingström, Jamil Yousef, Cecilia Hellström, My Hedhammar, Sebastian Havervall, Charlotte Thålin, Elisa Pin, … Sophia Hober *New Biotechnology* (2022-01) https://doi.org/gm9pkb DOI: 10.1016/j.nbt.2021.10.002 · PMID: 34628049 · PMCID: PMC8495044

112. **Two-Year Prospective Study of the Humoral Immune Response of Patients with Severe Acute Respiratory Syndrome** Wei Liu, Arnaud Fontanet, Pan-He Zhang, Lin Zhan, Zhong-Tao Xin, Laurence Baril, Fang Tang, Hui Lv, Wu-Chun Cao *The Journal of Infectious Diseases* (2006-03-15) https://doi.org/cmzn2k DOI: 10.1086/500469 · PMID: 16479513 · PMCID: PMC7109932

113. **The time course of the immune response to experimental coronavirus infection of man** KA Callow, HF Parry, M Sergeant, DAJ Tyrrell *Epidemiology and Infection* (2009-05-15) https://doi.org/c9pnmg DOI: 10.1017/s0950268800048019 · PMID: 2170159 · PMCID: PMC2271881

114. **Antibody Responses 8 Months after Asymptomatic or Mild SARS-CoV-2 Infection** Pyoeng Gyun Choe, Kye-Hyung Kim, Chang Kyung Kang, Hyeon Jeong Suh, EunKyo Kang, Sun Young Lee, Nam Joong Kim, Jongyoun Yi, Wan Beom Park, Myoung-don Oh *Emerging Infectious Diseases* (2021-03) https://doi.org/ghs9kq DOI: 10.3201/eid2703.204543 · PMID: 33350923

115. **Immunological memory to SARS-CoV-2 assessed for up to 8 months after infection** Jennifer M Dan, Jose Mateus, Yu Kato, Kathryn M Hastie, Esther Dawen Yu, Caterina E Faliti, Alba Grifoni, Sydney I Ramirez, Sonya Haupt, April Frazier, … Shane Crotty *Science* (2021-01-06) https://doi.org/ghrv9b DOI: 10.1126/science.abf4063 · PMID: 33408181

116. **Rapid generation of durable B cell memory to SARS-CoV-2 spike and nucleocapsid proteins in COVID-19 and convalescence.** Gemma E Hartley, Emily SJ Edwards, Pei M Aui, Nirupama Varese, Stephanie Stojanovic, James McMahon, Anton Y Peleg, Irene Boo, Heidi E Drummer, PMark Hogarth, … Menno C van Zelm *Science immunology* (2020-12-22) https://www.ncbi.nlm.nih.gov/pubmed/33443036 DOI: 10.1126/sciimmunol.abf8891 · PMID: 33443036

117. **Persistence of SARS-CoV-2-specific B and T cell responses in convalescent COVID-19 patients 6–8 months after the infection** Natalia Sherina, Antonio Piralla, Likun Du, Hui Wan, Makiko Kumagai-Braesch, Juni Andréll, Sten Braesch-Andersen, Irene Cassaniti, Elena Percivalle, Antonella Sarasini, … Qiang Pan-Hammarström *Med* (2021-03) https://doi.org/gh3xkz DOI: 10.1016/j.medj.2021.02.001 · PMID: 33589885 · PMCID: PMC7874960

118. **Evolution of antibody immunity to SARS-CoV-2** Christian Gaebler, Zijun Wang, Julio CC Lorenzi, Frauke Muecksch, Shlomo Finkin, Minami Tokuyama, Alice Cho, Mila Jankovic, Dennis Schaefer-Babajew, Thiago Y Oliveira, … Michel C Nussenzweig *Nature* (2021-01-18) https://doi.org/fq6k DOI: 10.1038/s41586-021-03207-w · PMID: 33461210 · PMCID: PMC8221082

119. **Robust neutralizing antibodies to SARS-CoV-2 infection persist for months** Ania Wajnberg, Fatima Amanat, Adolfo Firpo, Deena R Altman, Mark J Bailey, Mayce Mansour, Meagan McMahon, Philip Meade, Damodara Rao Mendu, Kimberly Muellers, … Carlos Cordon-Cardo *Science* (2020-12-04) https://doi.org/fgfs DOI: 10.1126/science.abd7728 · PMID: 33115920 · PMCID: PMC7810037

120. **Longitudinal observation and decline of neutralizing antibody responses in the three months following SARS-CoV-2 infection in humans** Jeffrey Seow, Carl Graham, Blair Merrick, Sam Acors, Suzanne Pickering, Kathryn JA Steel, Oliver Hemmings, Aoife O'Byrne, Neophytos Kouphou, Rui Pedro Galao, … Katie J Doores *Nature Microbiology* (2020-10-26) https://doi.org/fh7j DOI: 10.1038/s41564-020-00813-8 · PMID: 33106674 · PMCID: PMC7610833

121. **Evolution of immune responses to SARS-CoV-2 in mild-moderate COVID-19** Adam K Wheatley, Jennifer A Juno, Jing J Wang, Kevin J Selva, Arnold Reynaldi, Hyon-Xhi Tan, Wen Shi Lee, Kathleen M Wragg, Hannah G Kelly, Robyn Esterbauer, … Stephen J Kent *Nature*

*Communications* (2021-02-19) https://doi.org/gh9vd5 DOI: 10.1038/s41467-021-21444-5 · PMID: 33608522 · PMCID: PMC7896046

122. **COVID-19-neutralizing antibodies predict disease severity and survival** Wilfredo F Garcia-Beltran, Evan C Lam, Michael G Astudillo, Diane Yang, Tyler E Miller, Jared Feldman, Blake M Hauser, Timothy M Caradonna, Kiera L Clayton, Adam D Nitido, … Alejandro B Balazs *Cell* (2021-01) https://doi.org/gh9vdx DOI: 10.1016/j.cell.2020.12.015 · PMID: 33412089 · PMCID: PMC7837114

123. **Children develop robust and sustained cross-reactive spike-specific immune responses to SARS-CoV-2 infection** Alexander C Dowell, Megan S Butler, Elizabeth Jinks, Gokhan Tut, Tara Lancaster, Panagiota Sylla, Jusnara Begum, Rachel Bruton, Hayden Pearce, Kriti Verma, … Shamez Ladhani *Nature Immunology* (2021-12-22) https://doi.org/gnz6x6 DOI: 10.1038/s41590-021-01089-8 · PMID: 34937928 · PMCID: PMC8709786

124. **ABO blood group is involved in the quality of the specific immune response anti-SARS-CoV-2** Sergio Gil-Manso, Iria Miguens Blanco, Bruce Motyka, Anne Halpin, Rocío López-Esteban, Verónica Astrid Pérez-Fernández, Diego Carbonell, Luis Andrés López-Fernández, Lori West, Rafael Correa-Rocha, Marjorie Pion *Virulence* (2021-12-30) https://doi.org/gpphhv DOI: 10.1080/21505594.2021.2019959 · PMID: 34967260

125. **Loss of Bcl-6-Expressing T Follicular Helper Cells and Germinal Centers in COVID-19** Naoki Kaneko, Hsiao-Hsuan Kuo, Julie Boucau, Jocelyn R Farmer, Hugues Allard-Chamard, Vinay S Mahajan, Alicja Piechocka-Trocha, Kristina Lefteri, Matthew Osborn, Julia Bals, … Shiv Pillai *Cell* (2020-10) https://doi.org/gg9rdv DOI: 10.1016/j.cell.2020.08.025 · PMID: 32877699 · PMCID: PMC7437499

126. **Robust SARS-CoV-2-specific T-cell immunity is maintained at 6 months following primary infection** J Zuo, A Dowell, H Pearce, K Verma, HM Long, J Begum, F Aiano, Z Amin-Chowdhury, B Hallis, L Stapley, … P Moss *Cold Spring Harbor Laboratory* (2020-11-02) https://doi.org/ghhrps DOI: 10.1101/2020.11.01.362319

127. **Persistent Cellular Immunity to SARS-CoV-2 Infection** Gaëlle Breton, Pilar Mendoza, Thomas Hagglof, Thiago Y Oliveira, Dennis Schaefer-Babajew, Christian Gaebler, Martina Turroja, Arlene Hurley, Marina Caskey, Michel C Nussenzweig *Cold Spring Harbor Laboratory* (2020-12-09) https://doi.org/ghs9kk DOI: 10.1101/2020.12.08.416636 · PMID: 33330867 · PMCID: PMC7743071

128. **Targets of T Cell Responses to SARS-CoV-2 Coronavirus in Humans with COVID-19 Disease and Unexposed Individuals** Alba Grifoni, Daniela Weiskopf, Sydney I Ramirez, Jose Mateus, Jennifer M Dan, Carolyn Rydyznski Moderbacher, Stephen A Rawlings, Aaron Sutherland, Lakshmanane Premkumar, Ramesh S Jadi, … Alessandro Sette *Cell* (2020-06) https://doi.org/ggzxz2 DOI: 10.1016/j.cell.2020.05.015 · PMID: 32473127 · PMCID: PMC7237901

129. **COVID-19 testing turns to T cells** Cormac Sheridan *Nature Biotechnology* (2021-05) https://doi.org/gprdmb DOI: 10.1038/s41587-021-00920-9 · PMID: 33981082

130. **Coronavirus (COVID-19) Update: FDA Authorizes Adaptive Biotechnologies T-Detect COVID Test** Office of the Commissioner *FDA* (2021-03-09) https://www.fda.gov/news-events/press-announcements/coronavirus-covid-19-update-fda-authorizes-adaptive-biotechnologies-t-detect-covid-test

131. **Rethinking Covid-19 Test Sensitivity — A Strategy for Containment** Michael J Mina, Roy Parker, Daniel B Larremore *New England Journal of Medicine* (2020-11-26) https://doi.org/ghdg6n DOI: 10.1056/nejmp2025631 · PMID: 32997903

132. **The time to do serosurveys for COVID-19 is now** Rosanna W Peeling, Piero L Olliaro *The Lancet Respiratory Medicine* (2020-09) https://doi.org/gg559s DOI: 10.1016/s2213-2600(20)30313-1 · PMID: 32717209 · PMCID: PMC7380934

133. **Accurate point-of-care serology tests for COVID-19** Charles F Schuler, Carmen Gherasim, Kelly O'Shea, David M Manthei, Jesse Chen, Don Giacherio, Jonathan P Troost, James L Baldwin, James R Baker *PLOS ONE* (2021-03-16) https://doi.org/gpq4mz DOI: 10.1371/journal.pone.0248729 · PMID: 33725025 · PMCID: PMC7963097

134. **Neutralizing Antibodies Against Severe Acute Respiratory Syndrome Coronavirus 2 (SARS-CoV-2) Variants Induced by Natural Infection or Vaccination: A Systematic Review and Pooled Analysis** Xinhua Chen, Zhiyuan Chen, Andrew S Azman, Ruijia Sun, Wanying Lu, Nan Zheng, Jiaxin Zhou, Qianhui Wu, Xiaowei Deng, Zeyao Zhao, … Hongjie Yu *Clinical Infectious Diseases* (2021-07-24) https://doi.org/gmdht2 DOI: 10.1093/cid/ciab646 · PMID: 34302458 · PMCID: PMC9016754

135. **COVID-19 serosurveys for public health decision making** Manoj V Murhekar, Hannah Clapham *The Lancet Global Health* (2021-05) https://doi.org/gh7598 DOI: 10.1016/s2214-109x(21)00057-7 · PMID: 33705691 · PMCID: PMC8049585

136. **COVID-19 Serological Tests: How Well Do They Actually Perform?** Abdi Ghaffari, Robyn Meurant, Ali Ardakani *Diagnostics* (2020-07-04) https://doi.org/gg4h62 DOI: 10.3390/diagnostics10070453 · PMID: 32635444 · PMCID: PMC7400479

137. **Serological tests for COVID-19** Katherine Bond, Eloise Williams, Benjamin P Howden, Deborah A Williamson *Medical Journal of Australia* (2020-09-06) https://doi.org/gpqt86 DOI: 10.5694/mja2.50766 · PMID: 32892381

138. **A high-throughput multiplexed microfluidic device for COVID-19 serology assays** Roberto Rodriguez-Moncayo, Diana F Cedillo-Alcantar, Pablo E Guevara-Pantoja, Oriana G Chavez-Pineda, Jose A Hernandez-Ortiz, Josue U Amador-Hernandez, Gustavo Rojas-Velasco, Fausto Sanchez-Muñoz, Daniel Manzur-Sandoval, Luis D Patino-Lopez, … Jose L Garcia-Cordero *Lab on a Chip* (2021) https://doi.org/gpqt7j DOI: 10.1039/d0lc01068e · PMID: 33319882

139. **A systematic review of antibody mediated immunity to coronaviruses: antibody kinetics, correlates of protection, and association of antibody responses with severity of disease** Angkana T Huang, Bernardo Garcia-Carreras, Matt DT Hitchings, Bingyi Yang, Leah

Katzelnick, Susan M Rattigan, Brooke Borgert, Carlos Moreno, Benjamin D Solomon, Isabel Rodriguez-Barraquer, … Derek AT Cummings *Cold Spring Harbor Laboratory* (2020-04-17) https://doi.org/ggsfmz DOI: 10.1101/2020.04.14.20065771 · PMID: 32511434

140. **Antibodies, Immunity, and COVID-19** Brad Spellberg, Travis B Nielsen, Arturo Casadevall *JAMA Internal Medicine* (2021-04-01) https://doi.org/gpphhs DOI: 10.1001/jamainternmed.2020.7986 · PMID: 33231673 · PMCID: PMC8371694

141. **Protection and waning of natural and hybrid COVID-19 immunity** Yair Goldberg, Micha Mandel, Yinon M Bar-On, Omri Bodenheimer, Laurence Freedman, Nachman Ash, Sharon Alroy-Preis, Amit Huppert, Ron Milo *Cold Spring Harbor Laboratory* (2021-12-05) https://doi.org/g9rq DOI: 10.1101/2021.12.04.21267114

142. **Identification and Development of Therapeutics for COVID-19** Halie M Rando, Nils Wellhausen, Soumita Ghosh, Alexandra J Lee, Anna Ada Dattoli, Fengling Hu, James Brian Byrd, Diane N Rafizadeh, Ronan Lordan, Yanjun Qi, … Casey S Greene *mSystems* (2021-12-21) https://pubmed.ncbi.nlm.nih.gov/34726496/ DOI: 10.1128/msystems.00233-21

143. **Strong associations and moderate predictive value of early symptoms for SARS-CoV-2 test positivity among healthcare workers, the Netherlands, March 2020** Alma Tostmann, John Bradley, Teun Bousema, Wing-Kee Yiek, Minke Holwerda, Chantal Bleeker-Rovers, Jaap ten Oever, Corianne Meijer, Janette Rahamat-Langendoen, Joost Hopman, … Heiman Wertheim *Eurosurveillance* (2020-04-23) https://doi.org/ggthwx DOI: 10.2807/1560-7917.es.2020.25.16.2000508 · PMID: 32347200 · PMCID: PMC7189649

144. **Application and optimization of RT-PCR in diagnosis of SARS-CoV-2 infection** Xiaoshuai Ren, Yan Liu, Hongtao Chen, Wei Liu, Zhaowang Guo, Yaqin Zhang, Chaoqun Chen, Jianhui Zhou, Qiang Xiao, Guanmin Jiang, Hong Shan *Cold Spring Harbor Laboratory* (2020-02-27) https://doi.org/gpq4k9 DOI: 10.1101/2020.02.25.20027755

145. **Correlation of Chest CT and RT-PCR Testing for Coronavirus Disease 2019 (COVID-19) in China: A Report of 1014 Cases** Tao Ai, Zhenlu Yang, Hongyan Hou, Chenao Zhan, Chong Chen, Wenzhi Lv, Qian Tao, Ziyong Sun, Liming Xia *Radiology* (2020-08) https://doi.org/ggmw6p DOI: 10.1148/radiol.2020200642 · PMID: 32101510

146. **Performance of Radiologists in Differentiating COVID-19 from Non-COVID-19 Viral Pneumonia at Chest CT** Harrison X Bai, Ben Hsieh, Zeng Xiong, Kasey Halsey, Ji Whae Choi, Thi My Linh Tran, Ian Pan, Lin-Bo Shi, Dong-Cui Wang, Ji Mei, … Wei-Hua Liao *Radiology* (2020-08) https://doi.org/ggnqw4 DOI: 10.1148/radiol.2020200823 · PMID: 32155105

147. **Covid-19: automatic detection from X-ray images utilizing transfer learning with convolutional neural networks** Ioannis D Apostolopoulos, Tzani A Mpesiana *Physical and Engineering Sciences in Medicine* (2020-04-03) https://doi.org/ggs448 DOI: 10.1007/s13246-020-00865-4 · PMCID: PMC7118364

148. **Common pitfalls and recommendations for using machine learning to detect and prognosticate for COVID-19 using chest radiographs and CT scans** Michael Roberts, Derek

Driggs, Matthew Thorpe, Julian Gilbey, Michael Yeung, Stephan Ursprung, Angelica I Aviles-Rivero, Christian Etmann, Cathal McCague, … Carola-Bibiane Schönlieb *Nature Machine Intelligence* (2021-03) https://doi.org/gjkjvw DOI: 10.1038/s42256-021-00307-0

149. **Epidemiologic surveillance for controlling Covid-19 pandemic: types, challenges and implications** Nahla Khamis Ibrahim *Journal of Infection and Public Health* (2020-11) https://doi.org/gk7ghp DOI: 10.1016/j.jiph.2020.07.019 · PMID: 32855090 · PMCID: PMC7441991

150. **Wastewater and public health: the potential of wastewater surveillance for monitoring COVID-19** Kata Farkas, Luke S Hillary, Shelagh K Malham, James E McDonald, David L Jones *Current Opinion in Environmental Science & Health* (2020-10) https://doi.org/gg4tb6 DOI: 10.1016/j.coesh.2020.06.001 · PMID: 32835157 · PMCID: PMC7291992

151. **Healthcare Workers** CDC *Centers for Disease Control and Prevention* (2020-02-11) https://www.cdc.gov/coronavirus/2019-ncov/hcp/testing-overview.html

152. **Report 13: Estimating the number of infections and the impact of non-pharmaceutical interventions on COVID-19 in 11 European countries** S Flaxman, S Mishra, A Gandy, H Unwin, H Coupland, T Mellan, H Zhu, T Berah, J Eaton, P Perez Guzman, … S Bhatt *Imperial College London* (2020-03-30) https://doi.org/ggrbmf DOI: 10.25561/77731

153. **NY Forward: a guide to reopening New York & building back better** (2020-05-15) https://www.governor.ny.gov/sites/governor.ny.gov/files/atoms/files/NYForwardReopeningGuide.pdf

154. **Antigen tests for COVID-19** Yuta Kyosei, Sou Yamura, Mayuri Namba, Teruki Yoshimura, Satoshi Watabe, Etsuro Ito *Biophysics and Physicobiology* (2021) https://doi.org/gpp4wr DOI: 10.2142/biophysico.bppb-v18.004 · PMID: 33954080 · PMCID: PMC8049777

155. **Covid-19: Show us evidence for lifting restrictions, doctors tell Johnson** Adele Waters *BMJ* (2022-02-15) https://doi.org/gprm69 DOI: 10.1136/bmj.o383 · PMID: 35168994

156. **Exit strategies from lockdowns due to COVID-19: a scoping review** Madhavi Misra, Harsha Joshi, Rakesh Sarwal, Krishna D Rao *BMC Public Health* (2022-03-12) https://doi.org/gprm7c DOI: 10.1186/s12889-022-12845-2 · PMID: 35279102 · PMCID: PMC8917328

157. **Viral Cultures for Coronavirus Disease 2019 Infectivity Assessment: A Systematic Review** Tom Jefferson, Elisabeth A Spencer, Jon Brassey, Carl Heneghan *Clinical Infectious Diseases* (2020-12-03) https://doi.org/ghpwks DOI: 10.1093/cid/ciaa1764 · PMID: 33270107 · PMCID: PMC7799320

158. **SARS-CoV-2, SARS-CoV, and MERS-CoV viral load dynamics, duration of viral shedding, and infectiousness: a systematic review and meta-analysis** Muge Cevik, Matthew Tate, Ollie Lloyd, Alberto Enrico Maraolo, Jenna Schafers, Antonia Ho *The Lancet*

*Microbe* (2021-01) https://doi.org/ghk47x DOI: 10.1016/s2666-5247(20)30172-5 · PMID: 33521734 · PMCID: PMC7837230

159.    **Long-term SARS-CoV-2 RNA shedding and its temporal association to IgG seropositivity** Vineet Agarwal, AJ Venkatakrishnan, Arjun Puranik, Christian Kirkup, Agustin Lopez-Marquez, Douglas W Challener, Elitza S Theel, John C O'Horo, Matthew J Binnicker, Walter K Kremers, … Venky Soundararajan *Cell Death Discovery* (2020-12) https://doi.org/gprm6q DOI: 10.1038/s41420-020-00375-y · PMID: 33298894 · PMCID: PMC7709096

160.    **Trajectory of Viral RNA Load Among Persons With Incident SARS-CoV-2 G614 Infection (Wuhan Strain) in Association With COVID-19 Symptom Onset and Severity** Helen C Stankiewicz Karita, Tracy Q Dong, Christine Johnston, Kathleen M Neuzil, Michael K Paasche-Orlow, Patricia J Kissinger, Anna Bershteyn, Lorna E Thorpe, Meagan Deming, Angelica Kottkamp, … Elizabeth R Brown *JAMA Network Open* (2022-01-10) https://doi.org/gprm58 DOI: 10.1001/jamanetworkopen.2021.42796 · PMID: 35006245 · PMCID: PMC8749477

161.    **Covid-19: Tests on students are highly inaccurate, early findings show** Stephen Armstrong *BMJ* (2020-12-23) https://doi.org/gprm67 DOI: 10.1136/bmj.m4941 · PMID: 33361271

162.    **Covid-19: Controversial rapid test policy divides doctors and scientists** Zosia Kmietowicz *BMJ* (2021-01-12) https://doi.org/gprm68 DOI: 10.1136/bmj.n81 · PMID: 33436413

163.    **Assessment of the Analytical Sensitivity of 10 Lateral Flow Devices against the SARS-CoV-2 Omicron Variant** Joshua Deerain, Julian Druce, Thomas Tran, Mitchell Batty, Yano Yoga, Michael Fennell, Dominic E Dwyer, Jen Kok, Deborah A Williamson *Journal of Clinical Microbiology* (2022-02-16) https://doi.org/gnvpb8 DOI: 10.1128/jcm.02479-21 · PMID: 34936477 · PMCID: PMC8849215

164.    **Clarifying the evidence on SARS-CoV-2 antigen rapid tests in public health responses to COVID-19** Michael J Mina, Tim E Peto, Marta García-Fiñana, Malcolm G Semple, Iain E Buchan *The Lancet* (2021-04) https://doi.org/gnmbjd DOI: 10.1016/s0140-6736(21)00425-6 · PMID: 33609444 · PMCID: PMC8049601

165.    **Test sensitivity is secondary to frequency and turnaround time for COVID-19 screening** Daniel B Larremore, Bryan Wilder, Evan Lester, Soraya Shehata, James M Burke, James A Hay, Milind Tambe, Michael J Mina, Roy Parker *Science Advances* (2021-01) https://doi.org/ghs8s7 DOI: 10.1126/sciadv.abd5393 · PMID: 33219112 · PMCID: PMC7775777

166.    **CDC chief says coronavirus cases may be 10 times higher than reported** Washington Post https://www.washingtonpost.com/health/2020/06/25/coronavirus-cases-10-times-larger/

167.    **Liverpool coronavirus (COVID-19) community testing pilot: full evaluation report summary** GOV.UK https://www.gov.uk/government/publications/liverpool-coronavirus-covid-19-

community-testing-pilot-full-evaluation-report-summary/liverpool-coronavirus-covid-19-community-testing-pilot-full-evaluation-report-summary

168. **Covid-19: Government rolls out twice weekly rapid testing to all in England** Gareth Iacobucci *BMJ* (2021-04-06) https://doi.org/gprcnx DOI: 10.1136/bmj.n902 · PMID: 33824178

169. **A comparative study of COVID-19 responses in South Korea and Japan: political nexus triad and policy responses** M Jae Moon, Kohei Suzuki, Tae In Park, Kentaro Sakuwa *International Review of Administrative Sciences* (2021-03-18) https://doi.org/gprcn2 DOI: 10.1177/0020852321997552 · PMCID: PMC8685564

170. **All things equal? Heterogeneity in policy effectiveness against COVID-19 spread in chile** Magdalena Bennett *World Development* (2021-01) https://doi.org/gjggkh DOI: 10.1016/j.worlddev.2020.105208 · PMID: 32994662 · PMCID: PMC7513907

171. **UK ending Covid testing 'very worrying' as WHO chief warns pandemic 'isn't over'** The Independent (2022-03-11) https://www.independent.co.uk/news/health/who-covid-testing-anil-soni-b2032884.html

172. **UK scales back routine covid-19 surveillance** Jonathan Clarke, Thomas Beaney, Azeem Majeed *BMJ* (2022-03-04) https://doi.org/gprcnz DOI: 10.1136/bmj.o562 · PMID: 35246445

173. https://www.9news.com/article/news/health/coronavirus/local-groups-continue-push-for-covid-testing-and-vaccinations/73-bbcd8384-d96a-425e-aaeb-16ac9f36e581

174. **Utah will stop daily COVID case counts, close test sites in wind-down, Gov. Cox announces** The Salt Lake Tribune https://www.sltrib.com/news/2022/02/18/utah-will-stop-daily/

175. **Safety, tolerability and viral kinetics during SARS-CoV-2 human challenge in young adults** Ben Killingley, Alex J Mann, Mariya Kalinova, Alison Boyers, Niluka Goonawardane, Jie Zhou, Kate Lindsell, Samanjit S Hare, Jonathan Brown, Rebecca Frise, … Christopher Chiu *Nature Medicine* (2022-03-31) https://doi.org/gptbvp DOI: 10.1038/s41591-022-01780-9 · PMID: 35361992

176. **Considerations for the Safe Operation of Schools During the Coronavirus Pandemic** Ronan Lordan, Samantha Prior, Elizabeth Hennessy, Amruta Naik, Soumita Ghosh, Georgios K Paschos, Carsten Skarke, Kayla Barekat, Taylor Hollingsworth, Sydney Juska, … Tilo Grosser *Frontiers in Public Health* (2021-12-16) https://doi.org/gprq6z DOI: 10.3389/fpubh.2021.751451 · PMID: 34976917 · PMCID: PMC8716382

177. **SARS-CoV-2 infection and transmission in school settings during the second COVID-19 wave: a cross-sectional study, Berlin, Germany, November 2020** Stefanie Theuring, Marlene Thielecke, Welmoed van Loon, Franziska Hommes, Claudia Hülso, Annkathrin von der Haar, Jennifer Körner, Michael Schmidt, Falko Böhringer, Marcus A Mall, … *Eurosurveillance* (2021-08-26) https://doi.org/gprq6x DOI: 10.2807/1560-7917.es.2021.26.34.2100184 · PMID: 34448448 · PMCID: PMC8393892

178. **End to quarantines, contact tracing among changes in new Oregon guidance for schools, starting March 12** Elizabeth Miller *OPB* (2022-03-02) https://www.opb.org/article/2022/03/02/oregon-schools-guidance-mask-mandate-end/

179. **Palm Beach County public schools to stop COVID-19 contact tracing** WPTV (2022-03-11) https://www.wptv.com/news/education/palm-beach-county-public-schools-to-stop-covid-19-contact-tracing

180. **Covid News: C.D.C. Drops Contact Tracing Recommendation** Adeel Hassan *The New York Times* (2022-03-02) https://www.nytimes.com/live/2022/03/02/world/covid-19-tests-cases-vaccine

181. **Children and COVID-19: State-Level Data Report** American Academy of Pediatrics http://www.aap.org/en/pages/2019-novel-coronavirus-covid-19-infections/children-and-covid-19-state-level-data-report/

182. **A fifth of all US child Covid deaths occurred during Omicron surge** Melody Schreiber *The Guardian* (2022-03-11) https://www.theguardian.com/world/2022/mar/11/us-child-covid-deaths-omicron-surge

183. **Global, regional, and national minimum estimates of children affected by COVID-19-associated orphanhood and caregiver death, by age and family circumstance up to Oct 31, 2021: an updated modelling study** HJuliette T Unwin, Susan Hillis, Lucie Cluver, Seth Flaxman, Philip S Goldman, Alexander Butchart, Gretchen Bachman, Laura Rawlings, Christl A Donnelly, Oliver Ratmann, … Lorraine Sherr *The Lancet Child & Adolescent Health* (2022-04) https://doi.org/gpr264 DOI: 10.1016/s2352-4642(22)00005-0 · PMID: 35219404 · PMCID: PMC8872796

184. **Five Ways that COVID-19 Diagnostics Can Save Lives: Prioritizing Uses of Tests to Maximize Cost-Effectiveness** Tristan Reed, William Waites, David Manheim, Damien de Walque, Chiara Vallini, Roberta Gatti, Timothy B Hallett *World Bank* (2021-02-23) https://openknowledge.worldbank.org/handle/10986/35150